%% file: arXiv2015-main.tex
\documentclass[floatfix,prd,epsfig,nofootinbib,superscriptaddress,onecolumn,amssymb]{revtex4}
\usepackage[letterpaper,bindingoffset=0.0in,left=1in,right=1in,top=1in,bottom=1in,footskip=.25in]{geometry}

\usepackage{units}
\usepackage{slashed}
\usepackage{slashed}
\usepackage{graphicx,color}
\usepackage{epsfig}
\usepackage{subfigure}
\usepackage{epsfig}
\usepackage{dcolumn}
\usepackage{bm}
\usepackage{color}
\usepackage{enumitem}

\def\lsim{\mathrel{\rlap{\lower4pt\hbox{\hskip1pt$\sim$}}
    \raise1pt\hbox{$<$}}}         
\def\gsim{\mathrel{\rlap{\lower4pt\hbox{\hskip1pt$\sim$}}
    \raise1pt\hbox{$>$}}}         

\newcommand{\mjd}{\sc{Majorana Demonstrator}\rm}
\newcommand{\mj}{\sc{Majorana}\rm}
\begin{document}

\title{The COHERENT Experiment at the Spallation Neutron Source}

\input{authors.tex}

\begin{abstract}The COHERENT collaboration's primary objective is to
  measure coherent elastic neutrino-nucleus scattering (CEvNS) using
  the unique, high-quality source of tens-of-MeV neutrinos provided by
  the Spallation Neutron Source (SNS) at Oak Ridge National Laboratory
  (ORNL).  
 In spite of its large cross section, the CEvNS process has
  never been observed, due to tiny energies of the resulting nuclear
  recoils which are out of reach for standard neutrino detectors. The
  measurement of CEvNS has now become feasible, thanks to the
  development of ultra-sensitive technology for rare decay and
  weakly-interacting massive particle (dark matter) searches.  The
  CEvNS cross section is cleanly predicted in the standard model;
  hence its measurement provides a standard model test.  It is
  relevant for supernova physics and supernova-neutrino detection, and
  enables validation of dark-matter detector background and
  detector-response models.  In the long term, precision measurement
  of CEvNS will address questions of nuclear structure.
COHERENT will deploy multiple detector technologies in a phased approach:
 a 14-kg CsI[Na] scintillating crystal,
15 kg of p-type point-contact germanium detectors, and 100
kg of liquid xenon in a two-phase time projection chamber.  Following 
an extensive background measurement campaign, a location in the SNS basement has proven to be neutron-quiet and suitable for deployment of the COHERENT detector suite.  
The simultaneous
deployment of the three COHERENT detector subsystems will test the $N^2$ dependence of the
cross section and ensure an unambiguous discovery of CEvNS.   This
document describes concisely the COHERENT physics motivations,
sensitivity and
plans for measurements at the SNS to be accomplished on a four-year timescale. 

\end{abstract}

\maketitle

\def\etal{{\it et al.}}
\def\etalmacro{\etal\ (\macro)}

\thispagestyle{empty}

\normalsize
\input{background_Introduction.tex}

\input{Prop_Research_Methods.tex}

\input{timeline.tex}

\input{summary.tex}

\vspace{0.1in}
\noindent
\textbf{Acknowledgments}

We are grateful for logistical support and advice from 
SNS (a DOE Office of Science facility) and ORNL personnel. Much of the background measurement work was
done using ORNL SEED funds, as well as Sandia Laboratories Directed Research and Development (LDRD) and NA-22 support. The RED-100 xenon detector has been developed and constructed under the contract of
NRNU MEPhI with Ministry of Education and Science of Russian Federation No. 11.G34.31.0049 from October 19, 2011.
Xenon detector deployment is supported by ORNL LDRD funds.  We thank Pacific Northwest National Laboratory colleagues for CsI[Tl] detector
contributions and Triangle Universities Nuclear Laboratory for making resources for various detector components available.    COHERENT collaborators are supported by the U.S. Department of Energy Office of Science,  the National Science Foundation, and the Sloan Foundation.

\bibliographystyle{vitae}
\bibliography{refs}

\end{document}

%% file: authors.tex
\author{D.~Akimov}
\affiliation{SSC RF Institute for Theoretical and Experimental Physics of National Research Centre ``Kurchatov Institute'', Moscow, 117218, Russian Federation}
\affiliation{National Research Nuclear University MEPhI (Moscow Engineering Physics Institute), Moscow, 115409, Russian Federation}

\author{P.~An}
\affiliation{Triangle Universities Nuclear Laboratory, Durham, North Carolina, 27708, USA}

\author{C.~Awe}
\affiliation{Department of Physics, Duke University, Durham, NC 27708, USA}
\affiliation{Triangle Universities Nuclear Laboratory, Durham, North Carolina, 27708, USA}

\author{P.S.~Barbeau}
\affiliation{Department of Physics, Duke University, Durham, NC 27708, USA}
\affiliation{Triangle Universities Nuclear Laboratory, Durham, North Carolina, 27708, USA}

\author{P.~Barton}
\affiliation{Lawrence Berkeley National Laboratory, Berkeley, CA 94720, USA}

\author{B.~Becker}
\affiliation{Department of Physics and Astronomy, University of Tennessee, Knoxville, TN 37996, USA}

\author{V.~Belov}
\affiliation{SSC RF Institute for Theoretical and Experimental Physics of National Research Centre ``Kurchatov Institute'', Moscow, 117218, Russian Federation}
\affiliation{National Research Nuclear University MEPhI (Moscow Engineering Physics Institute), Moscow, 115409, Russian Federation}

\author{A.~Bolozdynya}
\affiliation{National Research Nuclear University MEPhI (Moscow Engineering Physics Institute), Moscow, 115409, Russian Federation}

\author{A.~Burenkov}
\affiliation{SSC RF Institute for Theoretical and Experimental Physics of National Research Centre ``Kurchatov Institute'', Moscow, 117218, Russian Federation}
\affiliation{National Research Nuclear University MEPhI (Moscow Engineering Physics Institute), Moscow, 115409, Russian Federation}

\author{B.~Cabrera-Palmer}
\affiliation{Sandia National Laboratories, Livermore, CA 94550, USA}

\author{J.I.~Collar}
\affiliation{Enrico Fermi Institute, Kavli Institute for Cosmological Physics and
Department of Physics, University of Chicago, Chicago, IL 60637, USA}

\author{R.J.~Cooper}
\affiliation{Lawrence Berkeley National Laboratory, Berkeley, CA 94720, USA}

\author{R.L.~Cooper}
\affiliation{Department of Physics, New Mexico State University, Las Cruces, NM 88003, USA}

\author{C.~Cuesta}
\affiliation{Department of Physics, University of Washington, Seattle, WA 98195, USA}

\author{D.~Dean}
\affiliation{Oak Ridge National Laboratory, Oak Ridge, TN 37831, USA}

\author{J.~Detwiler}
\affiliation{Department of Physics, University of Washington, Seattle, WA 98195, USA}

\author{A.G.~Dolgolenko}
\affiliation{SSC RF Institute for Theoretical and Experimental Physics of National Research Centre ``Kurchatov Institute'', Moscow, 117218, Russian Federation}

\author{Y.~Efremenko}
\affiliation{National Research Nuclear University MEPhI (Moscow Engineering Physics Institute), Moscow, 115409, Russian Federation}
\affiliation{Department of Physics and Astronomy, University of Tennessee, Knoxville, TN 37996, USA}

\author{S.R.~Elliott}
\affiliation{Los Alamos National Laboratory, Los Alamos, NM, USA, 87545, USA}

\author{A. Etenko}
\affiliation{National Research Centre ``Kurchatov Institute'', Moscow, 117218, Russian Federation}
\affiliation{National Research Nuclear University MEPhI (Moscow Engineering Physics Institute), Moscow, 115409, Russian Federation}

\author{N.~Fields}
\affiliation{Enrico Fermi Institute, Kavli Institute for Cosmological Physics and
Department of Physics, University of Chicago, Chicago, IL 60637, USA}

\author{W.~Fox}
\affiliation{Department of Physics, Indiana University, Bloomington, IN, 47405, USA}

\author{A.~Galindo-Uribarri}
\affiliation{Oak Ridge National Laboratory, Oak Ridge, TN 37831, USA}
\affiliation{Department of Physics and Astronomy, University of Tennessee, Knoxville, TN 37996, USA}

\author{M.~Green}
\affiliation{Physics Department, North Carolina State University, Raleigh, NC 27695, USA}

\author{M.~Heath}
\affiliation{Department of Physics, Indiana University, Bloomington, IN, 47405, USA}

\author{S.~Hedges}
\affiliation{Department of Physics, Duke University, Durham, NC 27708, USA}
\affiliation{Triangle Universities Nuclear Laboratory, Durham, North Carolina, 27708, USA}

\author{D.~Hornback}
\affiliation{Oak Ridge National Laboratory, Oak Ridge, TN 37831, USA}

\author{E.B.~Iverson}
\affiliation{Oak Ridge National Laboratory, Oak Ridge, TN 37831, USA}

\author{L.~Kaufman}
\affiliation{Department of Physics, Indiana University, Bloomington, IN, 47405, USA}

\author{S.R.~Klein}
\affiliation{Lawrence Berkeley National Laboratory, Berkeley, CA 94720, USA}

\author{A.~Khromov}
\affiliation{National Research Nuclear University MEPhI (Moscow Engineering Physics Institute), Moscow, 115409, Russian Federation}

\author{A.~Konovalov}
\affiliation{SSC RF Institute for Theoretical and Experimental Physics of National Research Centre ``Kurchatov Institute'', Moscow, 117218, Russian Federation}
\affiliation{National Research Nuclear University MEPhI (Moscow Engineering Physics Institute), Moscow, 115409, Russian Federation}

\author{A.~Kovalenko}
\affiliation{SSC RF Institute for Theoretical and Experimental Physics of National Research Centre ``Kurchatov Institute'', Moscow, 117218, Russian Federation}
\affiliation{National Research Nuclear University MEPhI (Moscow Engineering Physics Institute), Moscow, 115409, Russian Federation}

\author{A.~Kumpan}
\affiliation{National Research Nuclear University MEPhI (Moscow Engineering Physics Institute), Moscow, 115409, Russian Federation}

\author{C.~Leadbetter}
\affiliation{Triangle Universities Nuclear Laboratory, Durham, North Carolina, 27708, USA}

\author{L.~Li}
\affiliation{Department of Physics, Duke University, Durham, NC 27708, USA}
\affiliation{Triangle Universities Nuclear Laboratory, Durham, North Carolina, 27708, USA}

\author{W.~Lu}
\affiliation{Oak Ridge National Laboratory, Oak Ridge, TN 37831, USA}

\author{Y.~Melikyan}
\affiliation{National Research Nuclear University MEPhI (Moscow Engineering Physics Institute), Moscow, 115409, Russian Federation}

\author{D.~Markoff}
\affiliation{Physics Department, North Carolina Central University, Durham, North Carolina 27707, USA}
\affiliation{Triangle Universities Nuclear Laboratory, Durham, North Carolina, 27708, USA}

\author{K.~Miller}
\affiliation{Department of Physics, Duke University, Durham, NC 27708, USA}
\affiliation{Triangle Universities Nuclear Laboratory, Durham, North Carolina, 27708, USA}

\author{M.~Middlebrook}
\affiliation{Oak Ridge National Laboratory, Oak Ridge, TN 37831, USA}

\author{P.~Mueller}
\affiliation{Oak Ridge National Laboratory, Oak Ridge, TN 37831, USA}

\author{P.~Naumov}
\affiliation{National Research Nuclear University MEPhI (Moscow Engineering Physics Institute), Moscow, 115409, Russian Federation}

\author{J.~Newby}
\affiliation{Oak Ridge National Laboratory, Oak Ridge, TN 37831, USA}

\author{D.~Parno}
\affiliation{Department of Physics, University of Washington, Seattle, WA 98195, USA}

\author{S.~Penttila}
\affiliation{Oak Ridge National Laboratory, Oak Ridge, TN 37831, USA}

\author{G. Perumpilly}
\affiliation{Enrico Fermi Institute, Kavli Institute for Cosmological Physics and
Department of Physics, University of Chicago, Chicago, IL 60637, USA}

\author{D.~Radford}
\affiliation{Oak Ridge National Laboratory, Oak Ridge, TN 37831, USA}

\author{H.~Ray}
\affiliation{Department of Physics, University of Florida, Gainesville, FL 32611, USA}

\author{J.~Raybern}
\affiliation{Department of Physics, Duke University, Durham, NC 27708, USA}
\affiliation{Triangle Universities Nuclear Laboratory, Durham, North Carolina, 27708, USA}

\author{D.~Reyna}
\affiliation{Sandia National Laboratories, Livermore, CA 94550, USA}

\author{G.C.~Rich\footnote{Also at Department of Physics and Astronomy, University of North Carolina at Chapel Hill, Chapel Hill, NC 27599, USA}}
\affiliation{Triangle Universities Nuclear Laboratory, Durham, North Carolina, 27708, USA}

\author{D.~Rimal}
\affiliation{Department of Physics, University of Florida, Gainesville, FL 32611, USA}

\author{D.~Rudik}
\affiliation{SSC RF Institute for Theoretical and Experimental Physics of National Research Centre ``Kurchatov Institute'', Moscow, 117218, Russian Federation}
\affiliation{National Research Nuclear University MEPhI (Moscow Engineering Physics Institute), Moscow, 115409, Russian Federation}

\author{K. Scholberg\footnote{Corresponding author}}\email{schol@phy.duke.edu}
\affiliation{Department of Physics, Duke University, Durham, NC 27708, USA}

\author{B.~Scholz}
\affiliation{Enrico Fermi Institute, Kavli Institute for Cosmological Physics and
Department of Physics, University of Chicago, Chicago, IL 60637, USA}

\author{W.M.~Snow}
\affiliation{Department of Physics, Indiana University, Bloomington, IN, 47405, USA}

\author{V.~Sosnovtsev}
\affiliation{National Research Nuclear University MEPhI (Moscow Engineering Physics Institute), Moscow, 115409, Russian Federation}

\author{A.~Shakirov}
\affiliation{National Research Nuclear University MEPhI (Moscow Engineering Physics Institute), Moscow, 115409, Russian Federation}

\author{S.~Suchyta}
\affiliation{Department of Nuclear Engineering, University of California, Berkeley, CA, 94720, USA}

\author{B.~Suh}
\affiliation{Department of Physics, Duke University, Durham, NC 27708, USA}
\affiliation{Triangle Universities Nuclear Laboratory, Durham, North Carolina, 27708, USA}

\author{R.~Tayloe}
\affiliation{Department of Physics, Indiana University, Bloomington, IN, 47405, USA}

\author{R.T.~Thornton}
\affiliation{Department of Physics, Indiana University, Bloomington, IN, 47405, USA}

\author{I.~Tolstukhin}
\affiliation{National Research Nuclear University MEPhI (Moscow Engineering Physics Institute), Moscow, 115409, Russian Federation}

\author{K.~Vetter}
\affiliation{Department of Nuclear Engineering, University of California, Berkeley, CA, 94720, USA}
\affiliation{Lawrence Berkeley National Laboratory, Berkeley, CA 94720, USA}

\author{C.H.~Yu}
\affiliation{Oak Ridge National Laboratory, Oak Ridge, TN 37831, USA}

%% file: background_Introduction.tex
\section{Introduction}\label{sec:intro}

Coherent elastic neutrino-nucleus scattering (CEvNS\footnote{Note there exist a number of abbreviations for this process in the literature, e.g., CNS, CNNS, CENNS.  We favor a version with ``E'' for ``elastic'' to distinguish the process from inelastic coherent pion production, which is commonly confused with CEvNS by members of the high energy physics community.  We prefer to replace the first ``N'' with ``v'', for ``neutrino'', for two reasons: first, ``NN'' means ``nucleon-nucleon'' to many in the nuclear physics community.  Second, this disambiguates from CENNS, which is the name of an experimental collaboration.  Finally, the Roman letter ``v'' is less cumbersome than the Greek letter ``$\nu$''. }) was predicted 40 years ago as a consequence of the neutral weak current~\cite{PhysRevD.9.1389}. 
This standard-model process remains unobserved due to the daunting technical requirements: very low nuclear recoil energy thresholds, intense sources/large target masses, and low backgrounds. 
Employing state-of-the-art low-energy-threshold detector technology coupled with the intense stopped-pion neutrino source available
at the Spallation Neutron Source (SNS) at Oak Ridge National Laboratory (ORNL), 
the COHERENT Collaboration aims to measure CEvNS and to use it as a tool to search for physics beyond the standard model.
A suite of three detector subsystems (CsI[Na] scintillating crystals, p-type point-contact germanium detectors, and a two-phase liquid xenon (LXe) time projection chamber) will be deployed in the basement of the SNS, taking advantage of decades of detector development in the dark-matter direct-detection community.  
The COHERENT three-detector approach will enable unambiguous discovery of CEvNS.
The immediate experimental impacts of the search envisioned with this detector suite include:

\begin{itemize}
\item The first measurement of CEvNS and its predicted proportionality to neutron number squared, $N^2$.
\item A precision cross section measurement on multiple targets to test for non-standard neutrino interactions, for which the interaction depends on the quark makeup of the nucleus~\cite{Scholberg:2005qs,Barranco:2007tz}. 
\item Systematic characterization of low-threshold recoil detectors to validate experimental background and detector-response models, given that
CEvNS of natural neutrinos is an irreducible background for dark matter WIMP (Weakly Interacting Massive Particle) searches~\cite{billard:2014}.

\end{itemize}

Furthermore, the CEvNS process has one of the largest cross sections relevant for supernova dynamics and plays a significant role in core-collapse processes~\cite{Wilson:1974zz,Horowitz:2004pv}, and therefore should be measured to validate models of their behavior.   Dark matter detectors will also be able to detect CEvNS processes in the event of a nearby core-collapse supernova burst~\cite{Horowitz:2003cz,Chakraborty:2013zua}, which will be sensitive to the full flavor content of the supernova signal.

As a secondary goal, COHERENT will perform measurements of the charged-current cross sections Pb($\nu_e$,n), Fe($\nu_e$,n), and Cu($\nu_e$,n), which result in the emission of background-inducing fast neutrons.
The measurement of this cross section on lead has implications for supernova neutrino detection in the HALO supernova neutrino detector~\cite{Duba:2008zz}.
These ($\nu$, n) interactions may also influence nucleosynthesis in certain astrophysical environments~\cite{woosley1990:nuProcess,qian1997:NINSnucleosynthesis}.

	\section{Coherent Elastic Neutrino-Nucleus Scattering Physics}

The coherence of the CEvNS process results in an enhanced neutrino-nucleus cross section that is approximately proportional to $N^2$, the square of the number of neutrons in the nucleus, due to the small weak charge of the proton. The coherence condition, in which the neutrino scatters off all nucleons in a nucleus in phase, requires that the wavelength of the momentum transfer is larger than the size of the target nucleus. Full coherence requires low-energy neutrinos (typically $<$ 50~MeV for medium-$A$ nuclei); as a result, the experimental signature for the process is a difficult-to-detect keV to sub-keV nuclear recoil, depending on the nuclear mass and neutrino energy.

\subsection{Measurement of the $N^2$ Cross Section}

\begin{figure}[htb]
\begin{center}
\includegraphics[width=5in]{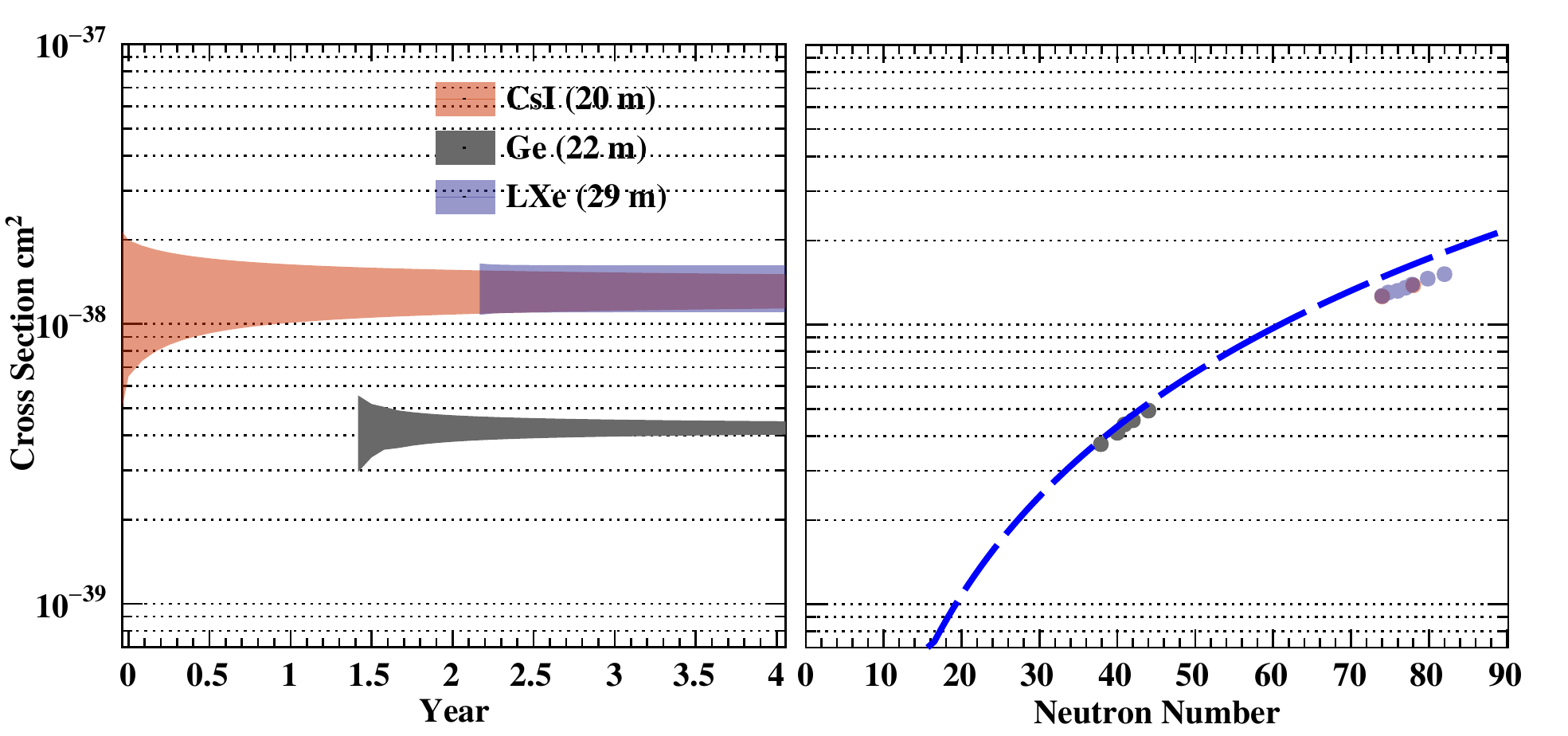}
\caption{\label{fig:cross-sections} Left:  Accounting for the deployment schedule, the expected precision for an $N^2$ cross-section measurement is shown for each subsystem. The dominant systematic uncertainties are due to imprecise knowledge of the nuclear recoil detector thresholds; the uncertainties on the rates due to quenching-factor uncertainties at threshold for each detector are taken to be: Ge, 2\% {\protect\cite{Barbeau:2007qi,Barbeau:2009zz}}; CsI[Na], 7\% {\protect\cite{cosinima}}; LXe, 13\% {\protect\cite{mock:2014NEST,szydagis:2015privateCom}}. The common $\sim$10\% neutrino flux uncertainty has been removed from this plot. Right:  Illustration of the $\propto N^2$ behavior of the CEvNS cross section versus neutron number $N$ (dashed blue line) for the relevant isotopes of COHERENT target materials. Deviations from the blue line are due to axial-vector currents on unpaired neutrons and protons and the increasing importance of the form factor for larger nuclei. }
\end{center}
\end{figure}

The cross section for CEvNS can be written as:
\begin{eqnarray}\label{eq:sevens}
\frac{d\sigma}{dT}_{coh} &=& \frac{G_F^2 M}{2\pi}\left[(G_V + G_A)^2 + (G_V - G_A)^2\left(1-\frac{T}{E_{\nu}}\right)^2 - (G_V^2 - G_A^2)\frac{MT}{E_{\nu}}\right] \\
G_V &=& (g_V^p Z + g_V^n N)F_{\rm nucl}^V(Q^2)\\
G_A &=& (g_A^p(Z_+ - Z_-) + g_A^n(N_+ - N_-))F_{\rm nucl}^A(Q^2),
\end{eqnarray}
where $G_F$ is the Fermi constant, $M$ is the nuclear mass, $T$ is the recoil energy, $E_{\nu}$  neutrino energy, 
$g_V^{n,p}$ and $g_A^{n,p}$ are vector and axial-vector coupling factors, respectively, for protons and neutrons, 
$Z$ and $N$ are the proton and neutron numbers, $Z_{\pm}$ and $N_{\pm}$ refer to the number of up or down nucleons, and $Q$ is the momentum transfer~\cite{Barranco:2005yy}.  The form factors $F_{\rm nucl}^A(Q^2)$ are point-like ($F(Q^2) = 1$) for interactions of low-energy neutrinos $<10$~MeV, but suppress the interaction rate as the wavelength of the momentum transfer becomes comparable to the size of the target nucleus (i.e.,~for higher neutrino energies and for heavier targets). The loss of coherence due to the nuclear form factor is expected to reduce the cross section for the SNS neutrinos by $<3$\% for the heavy xenon, cesium and iodine isotopes. 
The vector couplings appearing in $G_V$ and $G_A$ are written as:
\begin{eqnarray}
g_V^p &=& \rho_{\nu N}^{NC}\left(\frac{1}{2} - 2\hat{\kappa}_{\nu N}\sin^2 \theta_{W}\right) + 2\lambda^{uL} + 2\lambda^{uR} + \lambda^{dL} + \lambda^{dR} \\
g^n_V &=& -\frac{1}{2}\rho_{\nu N}^{NC} + \lambda^{uL}+\lambda^{uR} + 2\lambda^{dL} + 2\lambda^{dR},
\end{eqnarray}
where $\rho_{\nu N}^{NC}$, $\hat{\kappa}_{\nu N}$ are electroweak parameters,  $\lambda^{uL}, \lambda^{dL}, \lambda^{dR}, \lambda^{uR}$ are radiative corrections given in ~\cite{Barranco:2005yy,PDG2014}, and $\theta_W$ is the weak mixing angle.
The deployment of the COHERENT detector suite in an identified basement location at the SNS, which is $\sim$20 m from the source of neutrinos,
is expected to result in a clear observation of the coherent $N^2$ nature of the cross section within four years of operations at the SNS (Fig.~\ref{fig:cross-sections}). The expected precision of the cross section measurements, seen in the figure, quickly become dominated by the systematic uncertainty due to the imprecise knowledge of the nuclear recoil detector thresholds. This is especially true for the heavier Cs, I and Xe nuclei due to the lower average recoil energies for these species.

\subsection{Beyond-the-Standard-Model Physics Searches}

\begin{figure}[ht]
\centering
\subfigure[]{%
\includegraphics[width=0.45\linewidth]{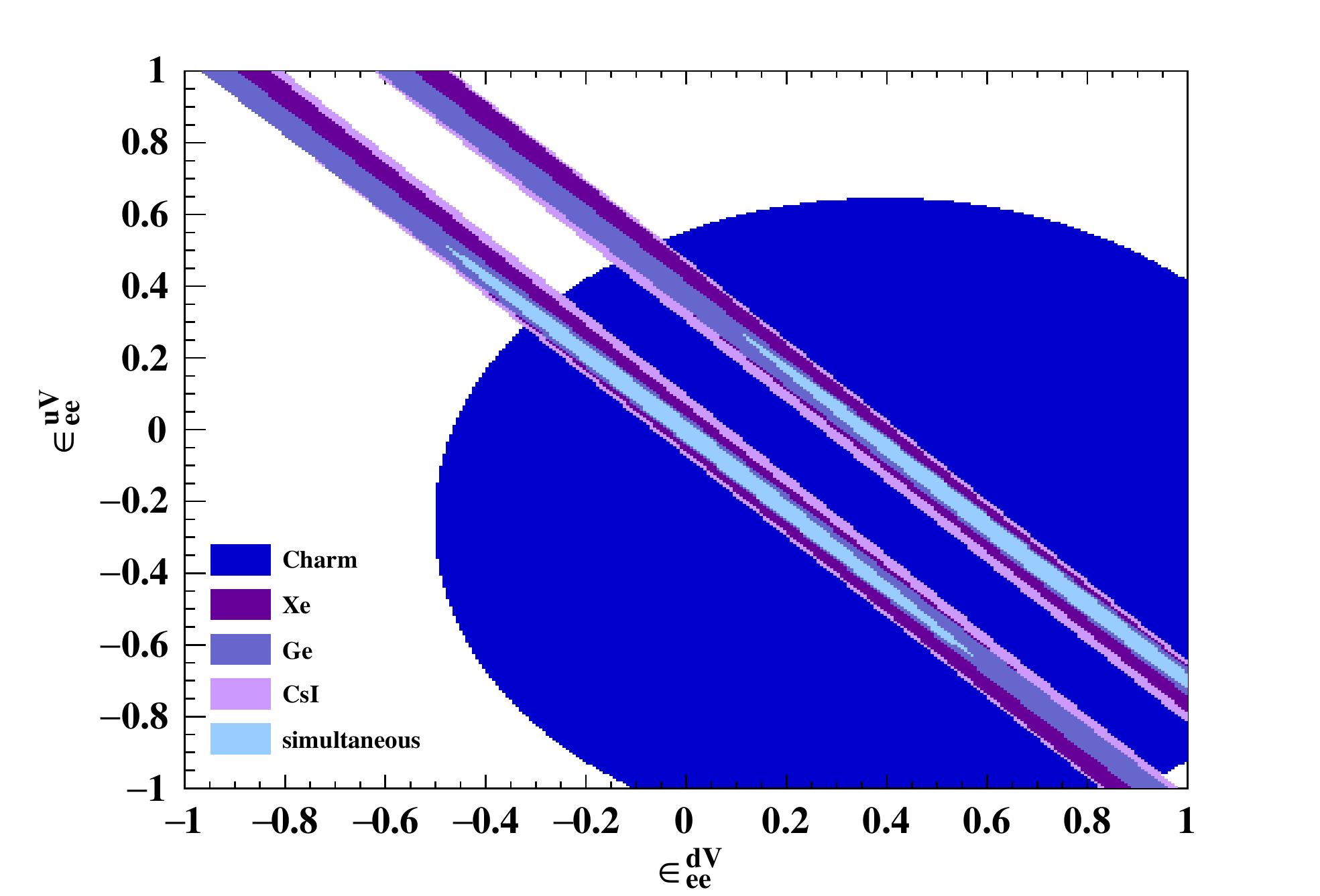}
}
\quad
\subfigure[]{%
\includegraphics[width=0.45\linewidth]{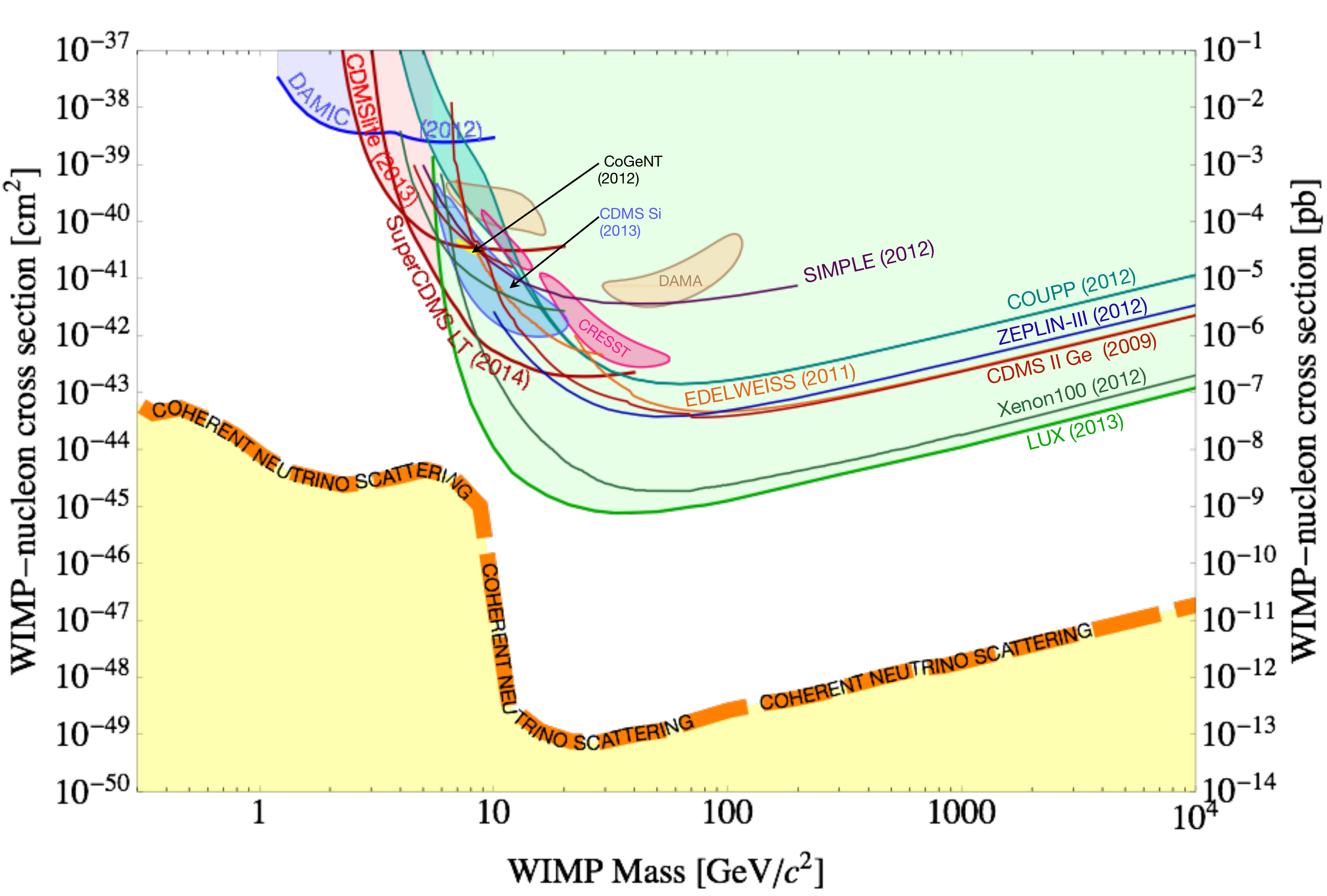}\label{fig:wimps}
}
\caption{\label{fig:NSI}(a) Allowed parameter space for two of the NSIs as constrained by the CHARM experiment {\protect\cite{Dorenbosch:1986tb}}; also shown is the predicted sensitivity obtained with the COHERENT detector suite. 
(b) WIMP dark-matter search parameter space, showing ``neutrino floor''~{\protect\cite{billard:2014}} from CEvNS of solar and atmospheric neutrinos as a thick orange dashed line.}
\end{figure}

Because the CEvNS cross section is cleanly predicted in the standard model, 
deviations can indicate new physics (e.g.,\cite{Barranco:2005yy, Barranco:2007tz, Harnik:2012ni,dutta2015}). For example, the CEvNS cross section for even-even nuclei, incorporating possible non-standard-interaction (NSI) neutral currents, can be parameterized as:
\begin{eqnarray}
\frac{d\sigma}{dT}_{coh} &=& \frac{G_f^2 M}{2\pi}G_V^2\left[1 + \left(1+\frac{T}{E_{\nu}}\right)^2 - \frac{MT}{E_{\nu}}\right]\\
G_V &=& ((g_V^p + 2\epsilon_{ee}^{uV} + \epsilon_{ee}^{dV})~Z + (g_V^n + \epsilon_{ee}^{uV} + 2\epsilon_{ee}^{dV}~)N)~F_{\rm nucl}^V
(Q^2),
\end{eqnarray}

\noindent
where the $\epsilon$'s represent new couplings~\cite{Scholberg:2005qs, Barranco:2005yy}.
Current neutrino-scattering constraints on the magnitude of non-zero values for $\epsilon_{ee}^{qV}$ from CHARM~\cite{Dorenbosch:1986tb} are of $\mathcal{O}(1)$ (see Fig.~\ref{fig:NSI}).  An improved search for NSIs can be performed by comparing measured CEvNS cross sections to standard-model expectations.  Assuming a systematic neutrino flux uncertainty of 10\%, the expected constraints from a null search for CsI[Na], Ge, Xe, and the combined analysis are shown in Fig.~\ref{fig:NSI} as diagonal bands. The angles vary slightly between the different isotopes due to different $N:Z$ ratios.  CEvNS using stopped-pion neutrinos can also constrain flavor-changing $\epsilon_{e\tau}$ NSI.   Sensitivity to NSI parameters can be improved with simultaneous measurements of the cross sections on different nuclei that factor out the neutrino flux uncertainty.

CEvNS measurements at a stopped-pion source can also probe beyond-the-standard-model physics by searching for new light weakly-coupled states produced in the proton target~\cite{deNiverville:2015mwa}.

\subsection{Relevance for Direct Dark Matter Detection Experiments}

CEvNS has long been closely linked to direct dark matter searches~\cite{Drukier:1983gj}.
The CEvNS of solar and atmospheric neutrinos, which produce single-scatter recoils identical to those expected from WIMPs, is recognized as an irreducible background for dark-matter WIMP searches for next-generation dark matter experiments~\cite{Monroe:2007xp,Gutlein:2010tq,cushman2013:snowmassDM,anderson:2011cevnsDM, billard:2014,Gutlein:2014gma}; see Fig.~\ref{fig:wimps}. Large dark-matter detectors may eventually be able to do solar neutrino physics using CEvNS~\cite{Billard:2014yka}.

The three detector technologies and materials proposed within the COHERENT program overlap well with those in use by the WIMP community.  Next-generation experiments such as LZ~\cite{malling:2011LZ},
XENON1T~\cite{Aprile:2012zx} and PandaX~\cite{Cao:2014jsa} utilize a liquid xenon time projection chamber;  SuperCDMS~\cite{superCDMS:2014PRL} uses germanium detectors; and the KIMS collaboration is conducting a WIMP search with CsI[Tl] crystals~\cite{kims:2012PRL}.
In these cases, by utilizing the intense SNS neutrino source, COHERENT can provide detector-specific response information for CEvNS interactions as a supplement to other low-energy calibration techniques. 
In addition to detector-response understanding, by measuring the CEvNS cross section on several different nuclei and observing the dependence of the interaction on proton and neutron numbers, the results obtained by COHERENT can serve to reduce systematic uncertainties associated with this cross section and its contribution to backgrounds in WIMP detectors.
A further constraint on cross-section uncertainty is afforded by the presence of some higher-energy ($\gsim 10$ MeV) neutrinos in the SNS beam: these neutrinos will provide information on nuclear form factors where deviation from unity may be present.

\subsection{Future Physics with CEvNS}\label{sec:future}

Beyond the scope of this proposal, a successful CEvNS program will enable new endeavors addressing further physics questions in the long term.

Of particular note is that 
the weak mixing angle ($\theta_{W}$) appears in the vector coupling factors of Eq.~\ref{eq:sevens}. 
It has long been predicted that a precision measurement of the CEvNS cross section would provide a sensitive test of the weak nuclear charge and of physics above the weak scale~\cite{Krauss:1991ba}.  Theoretical uncertainties are very small.
In addition to the terms in Eq.~\ref{eq:sevens}, there also exist axial-vector terms from strong quark and weak magnetism contributions for non-even-even nuclear targets that have larger theoretical uncertainties (as much as 5\% uncertainty for light nuclei). For the even-even targets we are considering, these uncertainties are $<$ 0.1\%.
While the total neutrino flux uncertainty is estimated to be $\sim$10\%,  improvements using multiple targets to a few-percent measurement of $\sin^2 \theta_W$ 
are conceivable.
CEvNS offers a measurement of the weak mixing angle at $Q\sim 40$ MeV/c,
where there is sensitivity in the weak mixing angle to models of dark Z bosons, which offer a possible explanation for the $(g-2)_{\mu}$ anomaly \cite{muonG-2:2006finalReport,jegerlehner:2009muonG-2} and a glimpse into dark sector physics \cite{davoudiasl2012:muonAnomaly,davoudiasl2012:darkZimplications,davoudiasl2014:darkBosons,kumar2013:lowEWMA}. 
A determination of $\sin^2 \theta_W$ produced by COHERENT (the first by this method) will have substantially larger uncertainty than those from parity-violating electron scattering experiments (see~\cite{erler2014:eScatReview} for a comprehensive review); however, this measurement will be subject to different systematic uncertainties and will provide an independent reckoning at comparable values of momentum transfer.

Additional examples of future CEvNS physics include:
\begin{itemize}
\item A measurement of the CEvNS interaction spectrum is sensitive to the magnitude of the neutrino magnetic moment~\cite{Dodd:1991ni,Scholberg:2005qs,Kosmas:2015sqa} resulting in a characteristic enhancement of low-energy recoils. Because the neutrino magnetic moment is predicted to be extremely tiny in the standard model, a measurement of non-zero magnetic moment would indicate new physics.

\item Precision measurement of the ratio of the muon and electron neutrino cross sections ($\sim$few~\%), which can be done by comparing prompt ($\nu_\mu$) and delayed $\nu_e$ stopped-pion signals, will be sensitive to the effective neutrino charge radius~\cite{Papavassiliou:2005cs}.  
\item Development of this flavor-blind neutral-current CEvNS detection capability will provide a natural tool to search for oscillations into sterile neutrinos in near and far detectors~\cite{Drukier:1983gj,Giunti:2006bj}.

\item The neutron distribution function (nuclear form factor) can be probed with measurements of the CEvNS recoil spectrum~\cite{Amanik:2009zz,Patton:2012jr}.
Incoherent contributions to the cross section are also of interest for study of nuclear axial structure~\cite{Moreno:2015bta}.

\end{itemize}

%% file: Prop_Research_Methods.tex
\section{The COHERENT Experiment}\label{sec:methods}

The COHERENT Collaboration has assembled a suite of detector
technologies suitable for the first observation of CEvNS at the SNS ---
the highest-flux pulsed, stopped-pion neutrino source currently
available. This effort leverages the technological development within
the dark-matter direct-detection community over the last decade by
deploying three mature low-threshold/low-background technologies capable of
observing low-energy nuclear recoils: CsI[Na] scintillator, p-type
point-contact (PPC) germanium detectors, and two-phase liquid xenon.
The combination of three separate detector subsystems (and four
elemental targets) with different systematics 
is vital for unambiguous measurement of CEvNS.
The use of targets with widely varying numbers of target neutrons
provides a ready test of the $N^2$ nature of the cross section, while
the use of two different technologies for the heaviest targets
(CsI[Na] and Xe) allows a convenient cross-check on the measured cross
section for the two systems most susceptible to systematic detector
response (quenching factor, QF) uncertainties.
The three detector subsystems will be installed in the SNS basement
hallway where a background measurement campaign (see
Sec.~\ref{sec:backgroundstudies}) has indicated very low beam-related
neutron backgrounds exist.  In this section we will describe the properties
of the SNS neutrino source, the results of background measurements, and the detectors.

\input{SNS_Nus.tex}

\input{Background_Studies.tex}

\input{Detectors.tex}

%% file: SNS_Nus.tex
\subsection{Neutrinos at the Spallation Neutron Source}

\begin{figure}[ht]
\centering
\subfigure[SNS neutrino energy spectrum]{%
\includegraphics[width=0.45\linewidth]{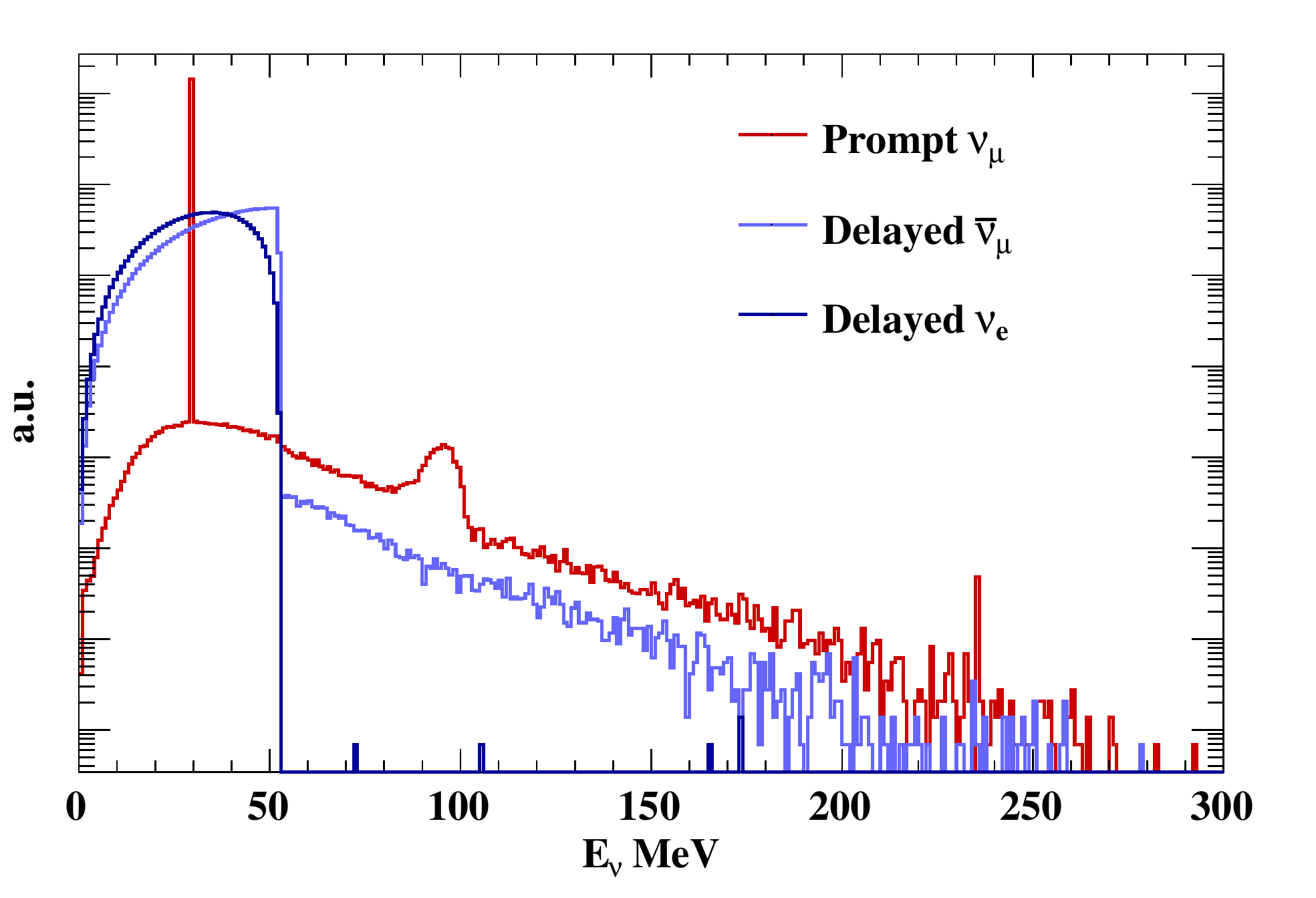}
\label{fig:snsisawesomea}}
\quad
\subfigure[SNS neutrino timing distribution]{%
\includegraphics[width=0.45\linewidth]{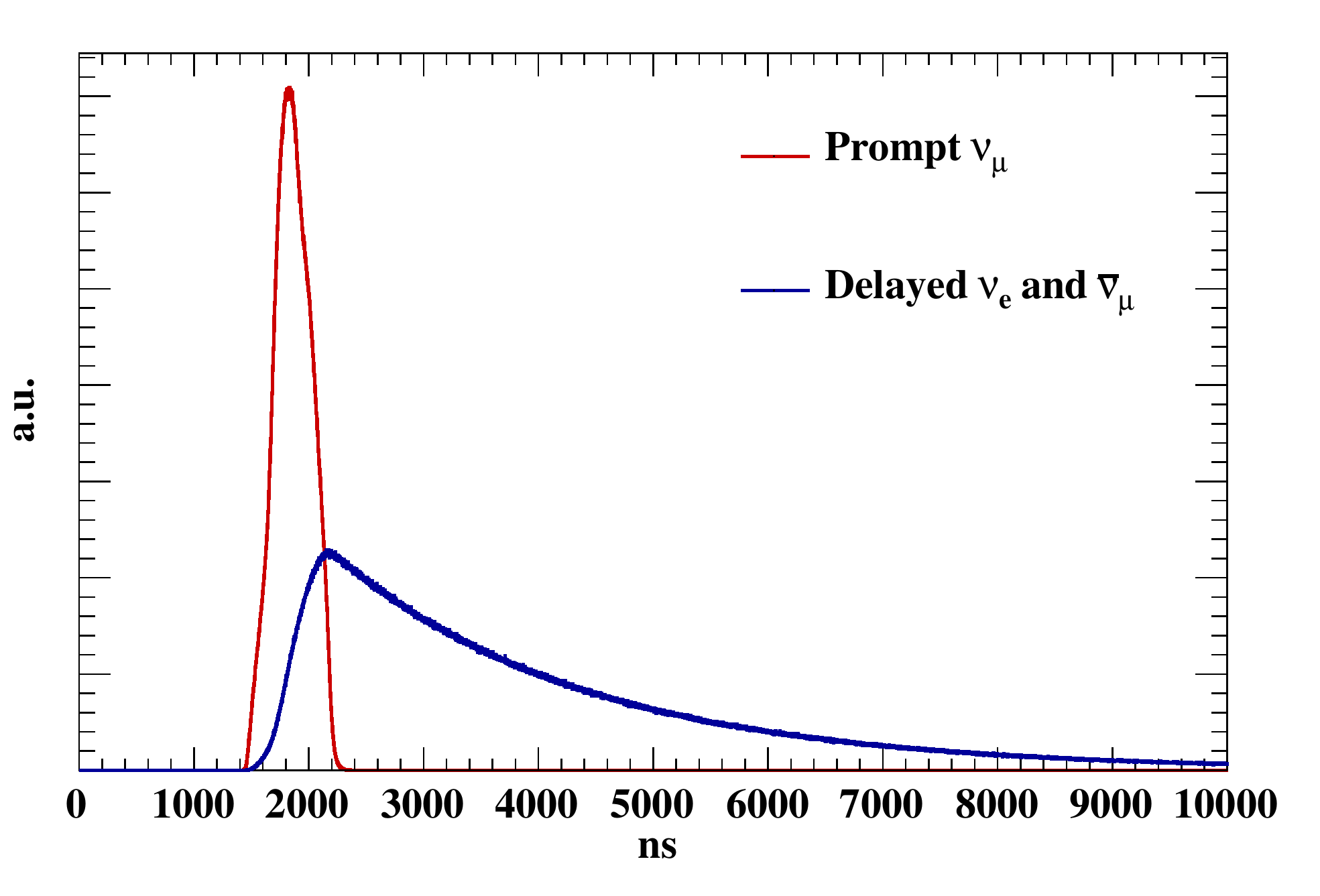}
\label{fig:snsisawesomeb}}
\caption{(a) Expected $\nu$ spectrum at the SNS, showing the very low level of decay-in-flight and other non-decay-at-rest flux, in arbitrary units; the integral is $4.3 \times 10^7$ neutrinos/cm$^2$/s at 20~m. (b) Time structure for prompt  and delayed neutrinos due to the 60-Hz pulses.}
\label{fig:snsisawesome}
\end{figure}

A stopped-pion beam has several advantages for CEvNS detection. First, the relatively high energies enhance the cross section ($\propto E^2$) while preserving coherence;  
cross sections at stopped-pion energies (up to 50~MeV) are about two orders of magnitude higher than at reactor energies ($\sim $3 MeV).   Second,
recoil energies (few to tens of keV) bring
detection of CEvNS within reach of the current generation of low-threshold
detectors. 
Finally, the different flavor content of the SNS flux ($\nu_\mu$, $\nu_e$ and $\bar{\nu}_\mu$) means 
that 
physics sensitivity is complementary to that for reactors ($\bar{\nu}_e$ only). See Fig.~\ref{fig:snsisawesome}.

The SNS produces an intense, isotropic stopped-pion neutrino flux, with a sharply pulsed timing structure that is highly beneficial for background rejection and precise characterization of the 
backgrounds not associated with the beam~\cite{Bolozdynya:2012xv}.  

A neutrino flux simulation has been developed for neutrino experiments at the SNS using GEANT4.10.1~\cite{Agostinelli:2002hh}.  Using this code, a beam of mono-energetic 1-GeV protons is simulated incident on a liquid mercury target.  
For the baseline simulation, a simplified version of the SNS target geometry was implemented.  
Neutrino spectra and flux are recorded for a COHERENT detector 
in the basement, 20 m from the target center at an angle of $-110^\circ$.
Resulting neutrino spectra and time distributions for different GEANT4 physics lists were compared; good agreement was found 
between ``QGSP\_BERT''  and ``QGSP\_INCLXX'' physics lists.  The flux prediction corresponding to QGSP\_BERT is used for the signal predictions shown in this proposal.

The results of the simulations show that the contribution to the neutrino spectrum from decay-in-flight and $\mu$-capture are expected to be very small (Fig. \ref{fig:snsisawesome}). The contribution to the CEvNS cross section from these high-energy neutrinos (E$>$50 MeV) is $<$1\%.  This contamination is at least two orders of magnitude smaller than at other existing facilities, e.g., at Fermilab~\cite{Brice:2013fwa} and J-PARC~\cite{Harada:2013yaa, Axani:2015dha}.  The SNS neutrino flux is also $\sim$80 times larger than that at the Fermilab Booster Neutrino Beam at the same distance from the source.

%% file: Background_Studies.tex
\subsection{Background Studies\label{sec:backgroundstudies}}

Understanding and reducing sources of background are critical goals of the COHERENT experimental program.  The short SNS duty cycle will reduce steady-state backgrounds due to radioactivity and cosmogenics by a factor of $10^3-10^4$. Steady-state backgrounds can also be understood using data taken outside the SNS beam window; however, beam-related backgrounds, especially fast neutrons, will be evaluated using ancillary measurements (described below) and modeling.  

\begin{figure}[ht]
\centering
\subfigure[]{%
\includegraphics[width=0.45\linewidth]{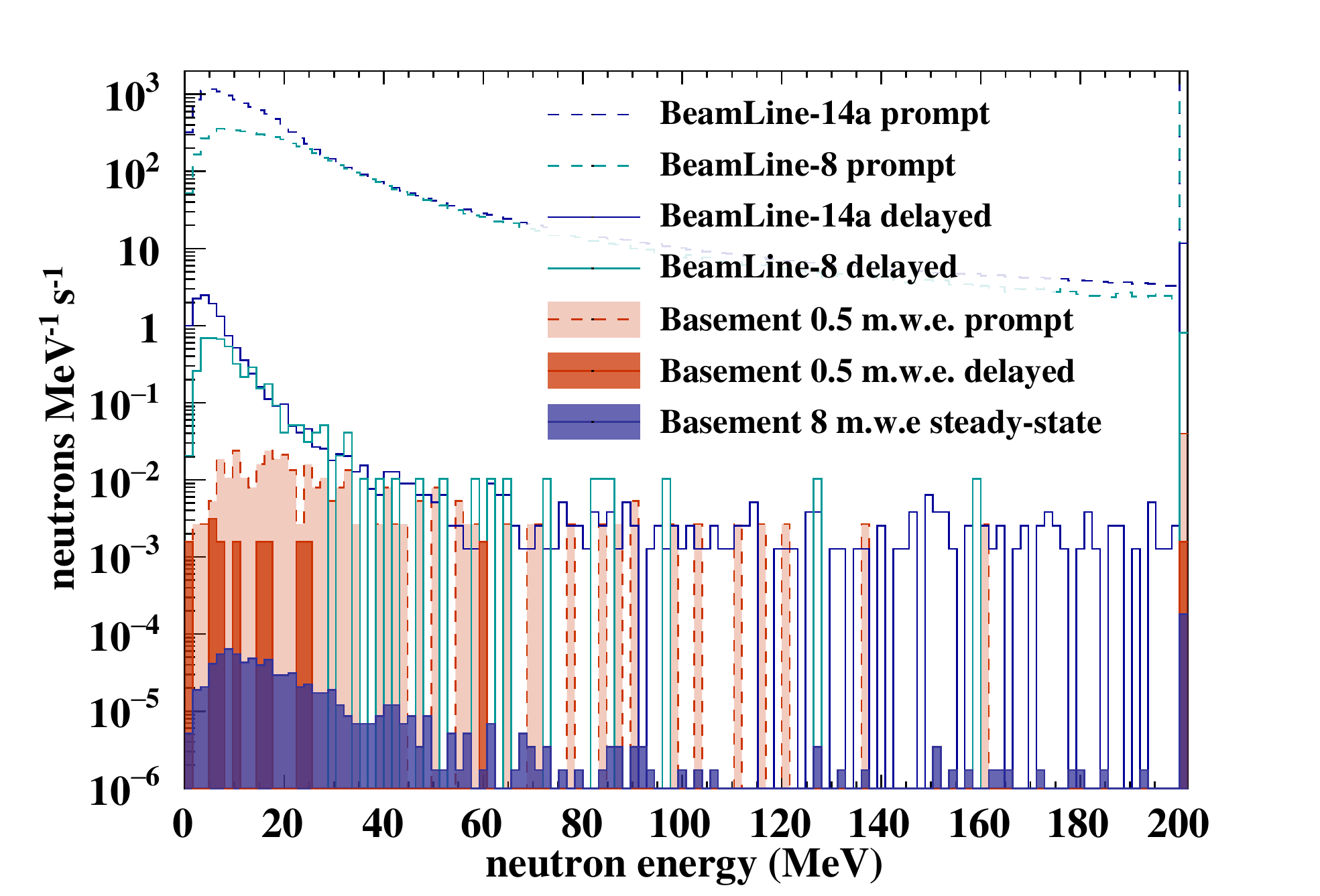}}
\quad
\subfigure[]{%
\includegraphics[width=0.45\linewidth]{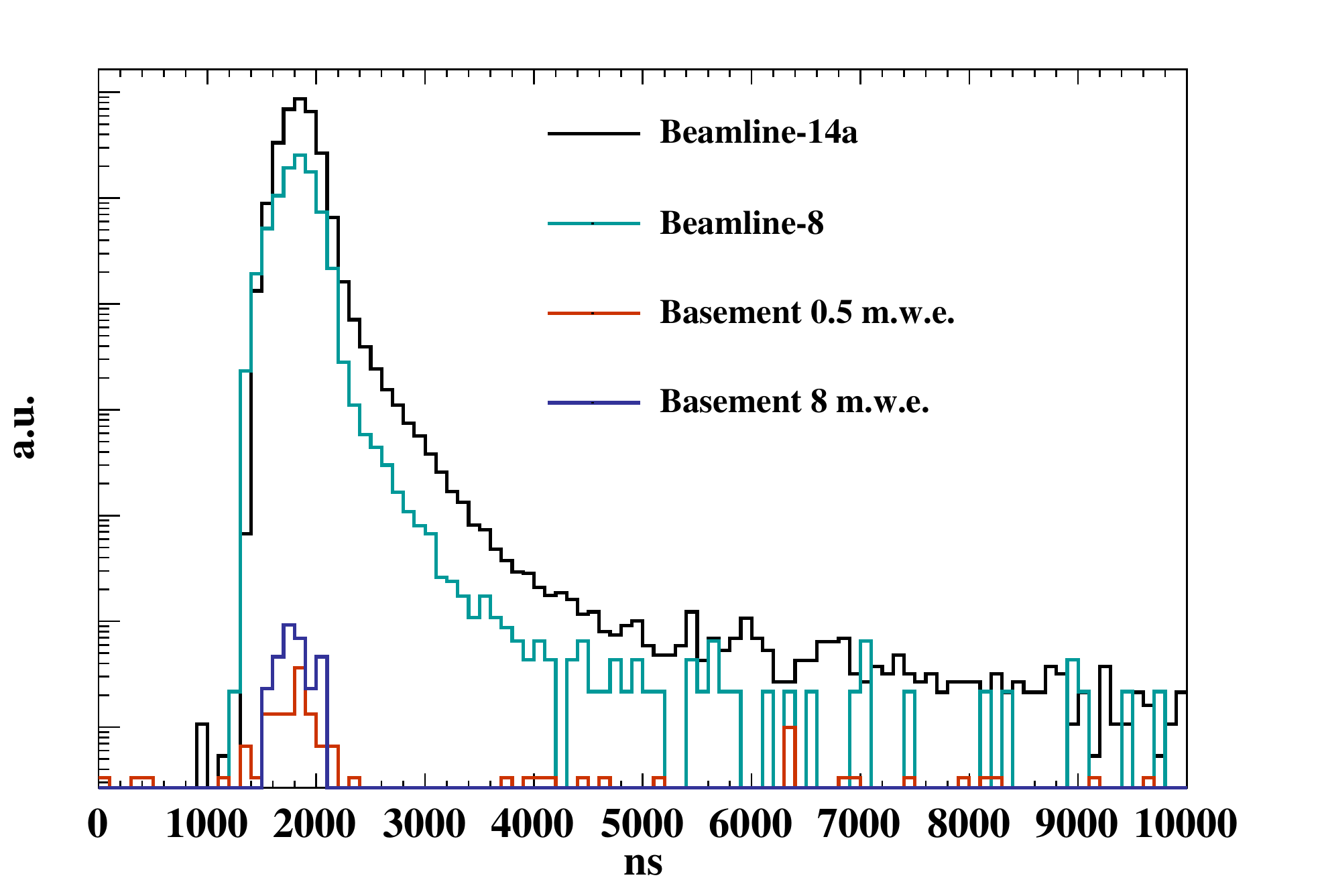}}
\caption{\label{fig:backgrounds_inbeam} (a)~Fast neutron spectra measured with the neutron scatter camera throughout the SNS facility.  A clear reduction by over four orders of magnitude from the experimental hall to the basement locations is seen. No neutron scatters were detected in the delayed window for the basement 8 m.w.e. location. (b) Arrival times of neutrons with respect to SNS beam timing signals. }
\end{figure}

A background measurement campaign has been underway since the fall of 2013.  Several detection systems have been used to measure beam-related neutron backgrounds, including: a single portable 5-liter liquid-scintillator detector to assess gross neutron rates at various locations within the SNS target hall and basement, a two-plane neutron scatter camera \cite{Brennan:2009} to provide detailed neutron spectra and some directional information (Fig.~\ref{fig:Nubes}(a)), and a single-plane liquid-scintillator array to provide systematic cross-checks of the neutron scatter camera data.  The measurements taken to date indicate that a basement location is very neutron-quiet; the direct beam-related neutrons are more than four orders of magnitude lower in the basement than on the experimental floor of the SNS (see Fig.~\ref{fig:backgrounds_inbeam}). 
The location is also protected from cosmic rays by $\sim$8 meters-water-equivalent (m.w.e.) of overburden.
The measured backgrounds in the basement are used to estimate the beam-related backgrounds for all three detector subsystems.

It is possible to effectively eliminate beam-related neutron backgrounds by taking advantage of the timing characteristics of the SNS (illustrated in Fig. \ref{fig:snsisawesome}).  The SNS provides beam timing signals that allow precise selection around the $\lsim$800-ns-width arrival time of protons on target. While a small background contribution from beam-related neutrons is expected to add to the $\nu_{\mu}$ signal in the ``prompt'' window (coincident with protons on target), measurements indicate that there are negligible beam-related neutrons expected in the ``delayed'' window, when the $\nu_{e}$ and $\bar{\nu}_{\mu}$ from muon decay arrive.

Rough acceptance and efficiency corrections to the measured neutron scatter camera spectra were performed in order to provide useful background estimates for the detector subsystem simulations (Fig.~\ref{fig:backgrounds_inbeam}). The flux measurements are dominated by uncertainties due to variations in the angular acceptance of the neutron scatter camera (factors of 3--5). Nevertheless, the measurements provide sufficient confidence that the experiment will succeed in the selected experimental locations, especially when signal analysis efforts are restricted to the delayed timing window.

To increase confidence in the understanding of the neutron background energy spectra, additional measurements will be made.
The neutron scatter camera will be positioned at a specific location near the expected CsI and Ge detectors, and
a significantly larger sample of data will be recorded, which will permit a refinement of the measured background contributions in the delayed window. In addition, ancillary measurements are planned with the complementary SciBath detector~\cite{Tayloe:2006ct, Cooper:2011kx,Brice:2013fwa}, a liquid-scintillator tracking detector, deployed a few meters from the neutron scatter camera for a simultaneous measurement. The combination of the two provides a systematic cross-check, increasing confidence in the understanding of the neutron background energy spectra.

Ongoing monitoring of the beam-related neutron backgrounds will be provided during the life of the experiment; however, the precision provided by either the neutron scatter camera or the SciBath detector is likely not required for the entire time.   We expect that single-channel liquid-scintillator cells will be adequate for continued monitoring.

	\subsubsection{Neutrino-Induced Neutron Backgrounds}
	The high-energy neutrinos from pion decay at rest have energies above the neutron separation threshold in $^{208}$Pb, a ubiquitous material in the detector radiological shields. The  charged-current interaction ($^{208}$Pb($\nu_e$, e)$^{208}$Bi), with subsequent prompt neutron emission, may produce significant numbers of \emph{background-producing} neutrons in the Pb shields, pulsed in time with the beam and sharing the 2.2-$\mu$s characteristic time-structure of the $\nu_e$'s due to the muon lifetime.  Other isotopes of Pb should have similarly  large neutron-ejection cross sections, and other elements commonly used for shielding (Fe, Cu) may also produce neutrino-induced neutrons (NINs).
	
	 As the exact cross sections are unknown (estimated theoretically within a factor of $\sim$3~\cite{Fuller:1998kb,Kolbe:2000np,Engel:2002hg}), a careful measurement of the production cross section is of great importance for background predictions. The measurement of these cross sections also has an impact on the ongoing HALO supernova neutrino detection experiment~\cite{Duba:2008zz, Vaananen:2011bf}. The spallation of neutrons from heavy elements is also expected to influence the nucleosynthesis of heavy elements in supernovae~\cite{qian1997:NINSnucleosynthesis,woosley1990:nuProcess}.

\begin{figure}[ht]
\centering
\subfigure[]{%
\includegraphics[width=0.24\linewidth]{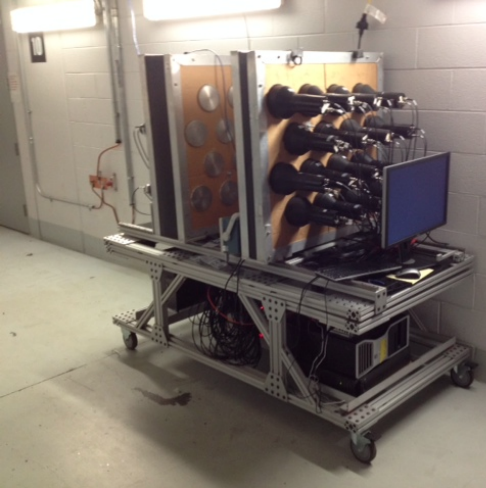}
}
\subfigure[]{%
\includegraphics[width=0.48\linewidth]{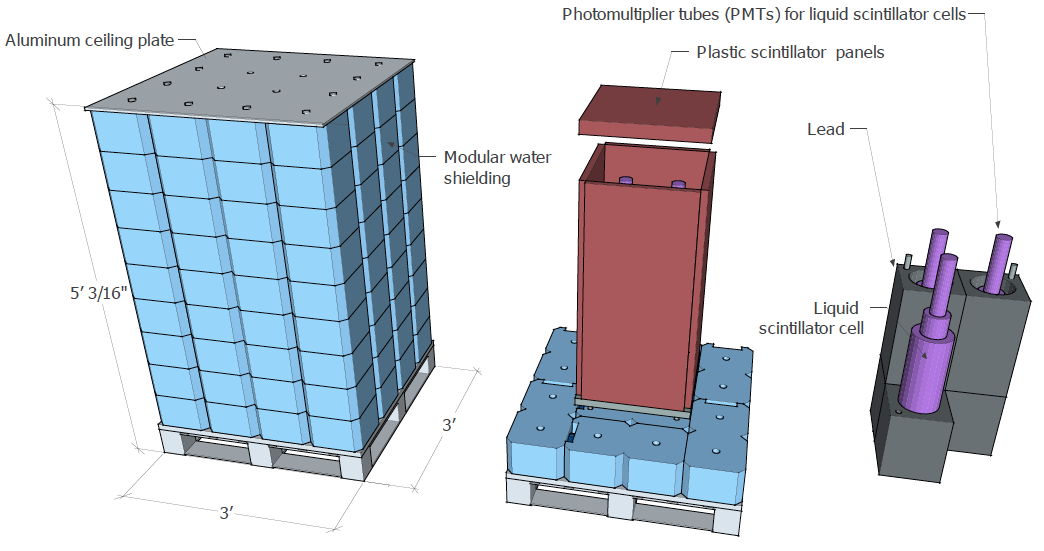}
}
\subfigure[]{%
\includegraphics[width=0.24\linewidth]{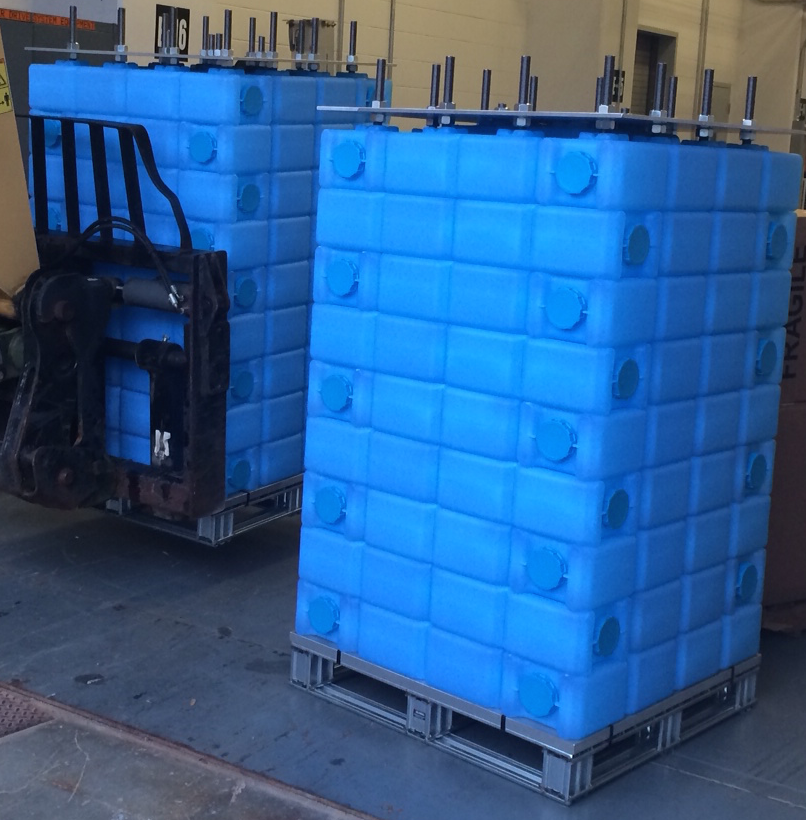}
}
\caption{\label{fig:Nubes} (a) The neutron scatter camera deployed in the SNS basement. (b) Schematic drawing of the detectors to measure the neutrino-induced neutron cross section on Pb, Fe, and Cu. (c) The recent arrival at the SNS building of the imperfectly-named ``Neutrino Cubes,'' modular neutrino-induced neutron experiments.}
\end{figure}

Dedicated apparatus containing liquid-scintillator detectors surrounded by lead or another target material and further surrounded by a muon veto and neutron moderator were designed and deployed to the SNS basement in September 2014 (see Figure~\ref{fig:Nubes}). These detector systems are expected to continue operation through the lifetime of the experiment.
The Collaboration will use these to measure the production cross sections of NINs in lead, iron, and copper at the SNS, both to evaluate the NIN background for CEvNS and as independent physics measurements.

%% file: Detectors.tex
\subsection{Detector Subsystems} 
After an extensive review of available detector technologies, the COHERENT Collaboration has selected three detector subsystems, each containing different target nuclei (summarized in Tab.~\ref{tab:detectors}).  The timing resolution of all three detector subsystems is sufficient to allow the observation of the characteristic 2.2-$\mu$s lifetime of muon-decay neutrinos, a further cross-check that any interactions are due to neutrinos from the SNS. The technologies are mature as all three are presently used for direct dark-matter detection experiments.

The Collaboration has already acquired some components of the COHERENT detector suite.
The CsI[Na] detector has already been
installed at the SNS.   The two-phase liquid xenon detector has already been constructed in Russia.  A total of 10~kg of the proposed 15~kg of Ge detectors are already available to the collaboration.  

\begin{figure}[ht]
\centering
\includegraphics[height=3.5in]{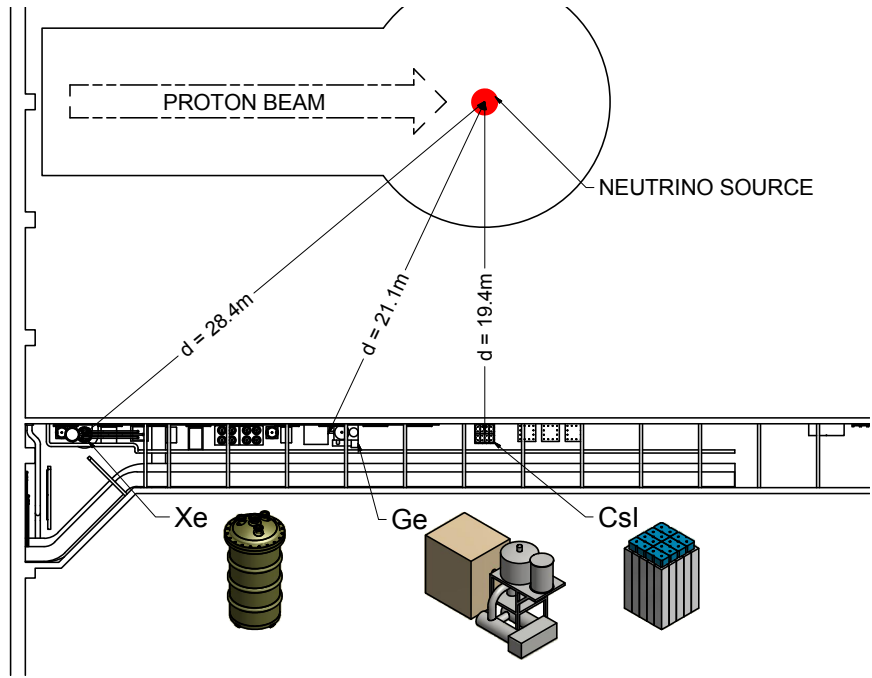}
\caption{Proposed siting in the SNS basement hallway.}\label{fig:neutrino_corridor}
\end{figure}

\begin{figure}[ht]
\centering
\subfigure[Recoil Spectra]{%
\includegraphics[height=3.0in]{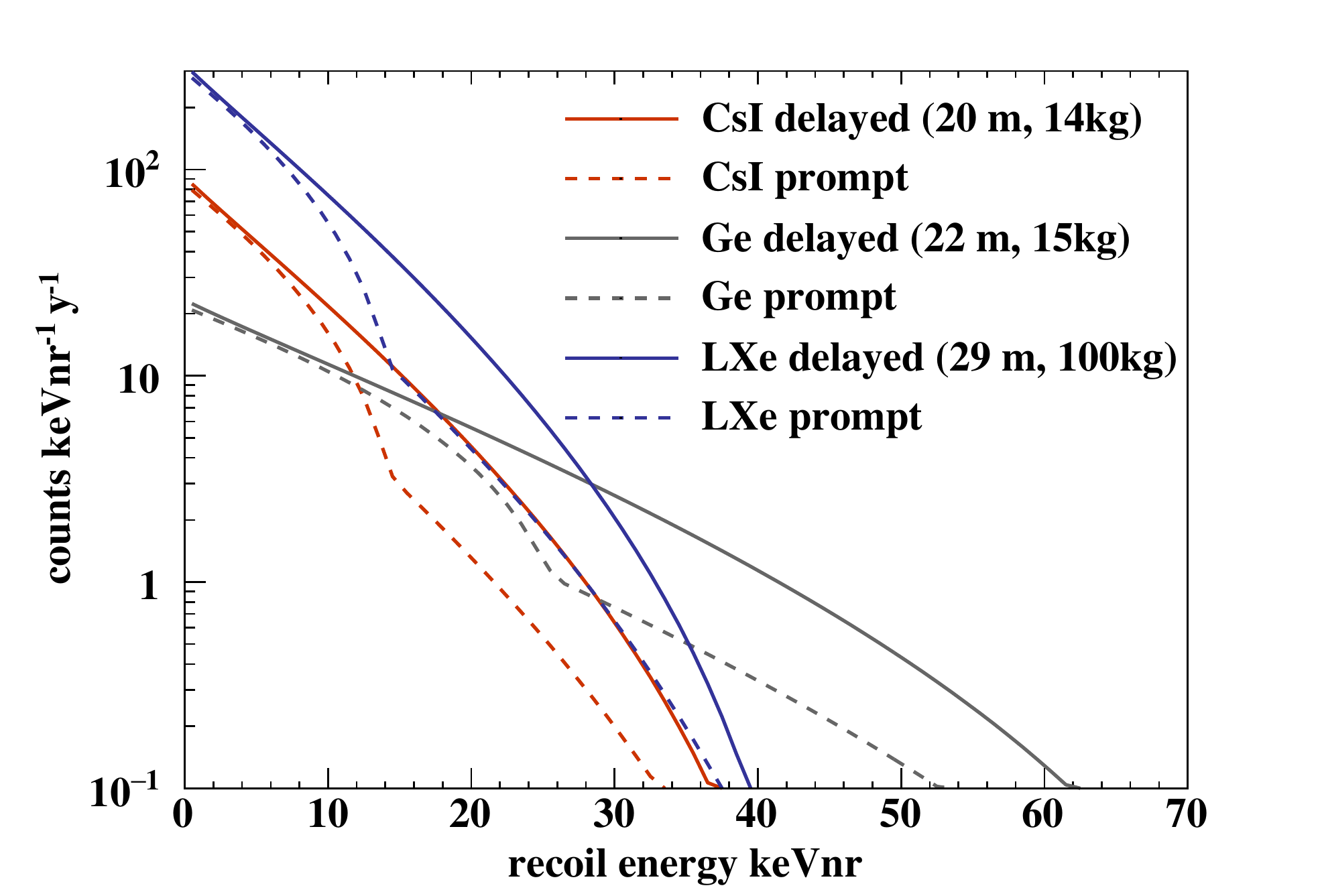}
\label{fig:RecoilRates}}
\caption{Differential recoil rates for the COHERENT suite of detectors in the SNS basement. Prompt $\nu_{\mu}$ neutrinos (arriving within 1 $\mu$s) have a small contamination ($\sim$20\%) from $\nu_e$ and $\bar{\nu}_{\mu}$.}
\label{fig:spectrum}
\end{figure}

All three detector subsystems will be deployed inside optimized shielding in the SNS basement location indicated to be suitable by the background measurement campaign described in Sec.~\ref{sec:backgroundstudies}.
Fig.~\ref{fig:neutrino_corridor} shows the proposed siting in the basement hallway.  The expected differential recoil rates for the proposed target masses and distances are shown in Fig.~\ref{fig:spectrum}.

\begin{table*}[htbp]
	\centering
 	\begin{tabular}{c|c|c|c|c|c}
		\hline
 		Nuclear& Technology & Mass & Dist. from & Recoil & CEvNS \\

		target & & (kg) &  source (m) & thresh. (keVnr)& detection goal
		\\ \hline 

		CsI[Na] & Scint. crystal & 14 & 20 & 5 & 5$\sigma$ in 2.0 yr run-time\\ 
		Ge & HPGe PPC & 15 & 22 & 5 & 5$\sigma$ in 0.5 yr run-time \\
		Xe & 2-phase Xe TPC & 100 & 29 & 4 & 5$\sigma$ in 0.2 yr run-time\\ 
		\hline
	\end{tabular}
	\caption{\label{tab:detectors}Parameters for the three COHERENT detector subsystems.}
 \end{table*}

\begin{figure}[ht]
\centering
\includegraphics[height=2.5in]{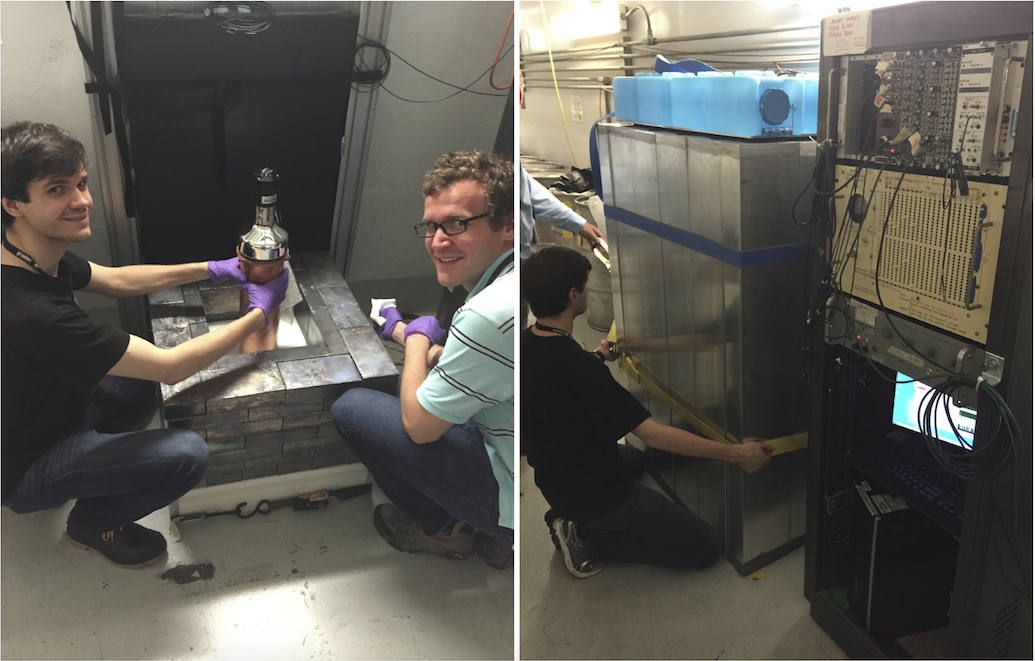}
\caption{The recent installation of the 14-kg, low-background CsI[Na] at the SNS. Successive layers of shielding include (inside-to-out): 7.62 cm HDPE, 15 cm of lead, a 5-cm thick muon veto and 10 cm of water neutron moderator.}\label{fig:csi_installation}
\end{figure}

\input{CsI.tex}

\begin{figure}[ht]
\centering
\subfigure[CsI]{%
\includegraphics[height=2.0in]{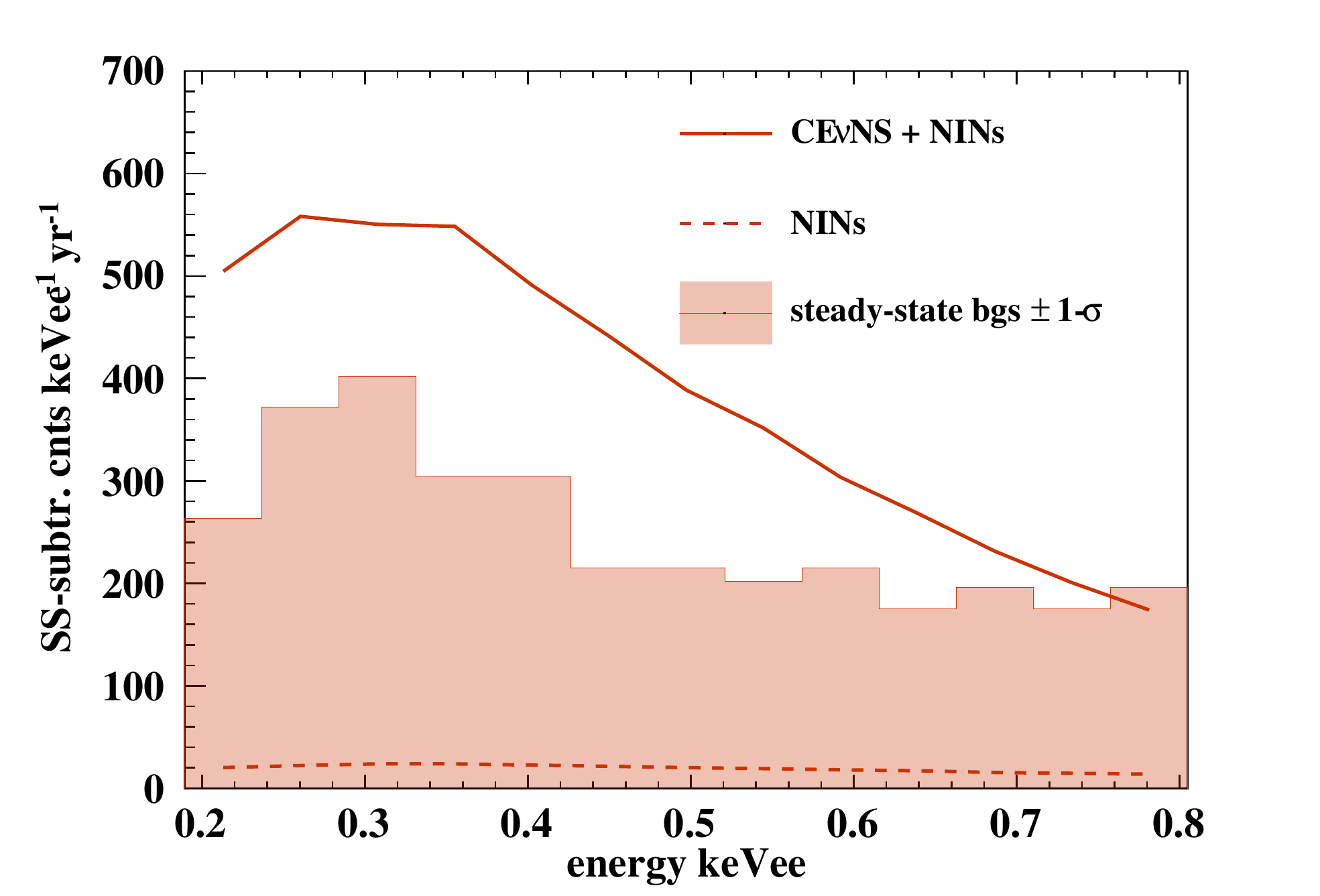}
\label{fig:spectrumcsi}}
\subfigure[Ge]{%
\includegraphics[height=2.0in]{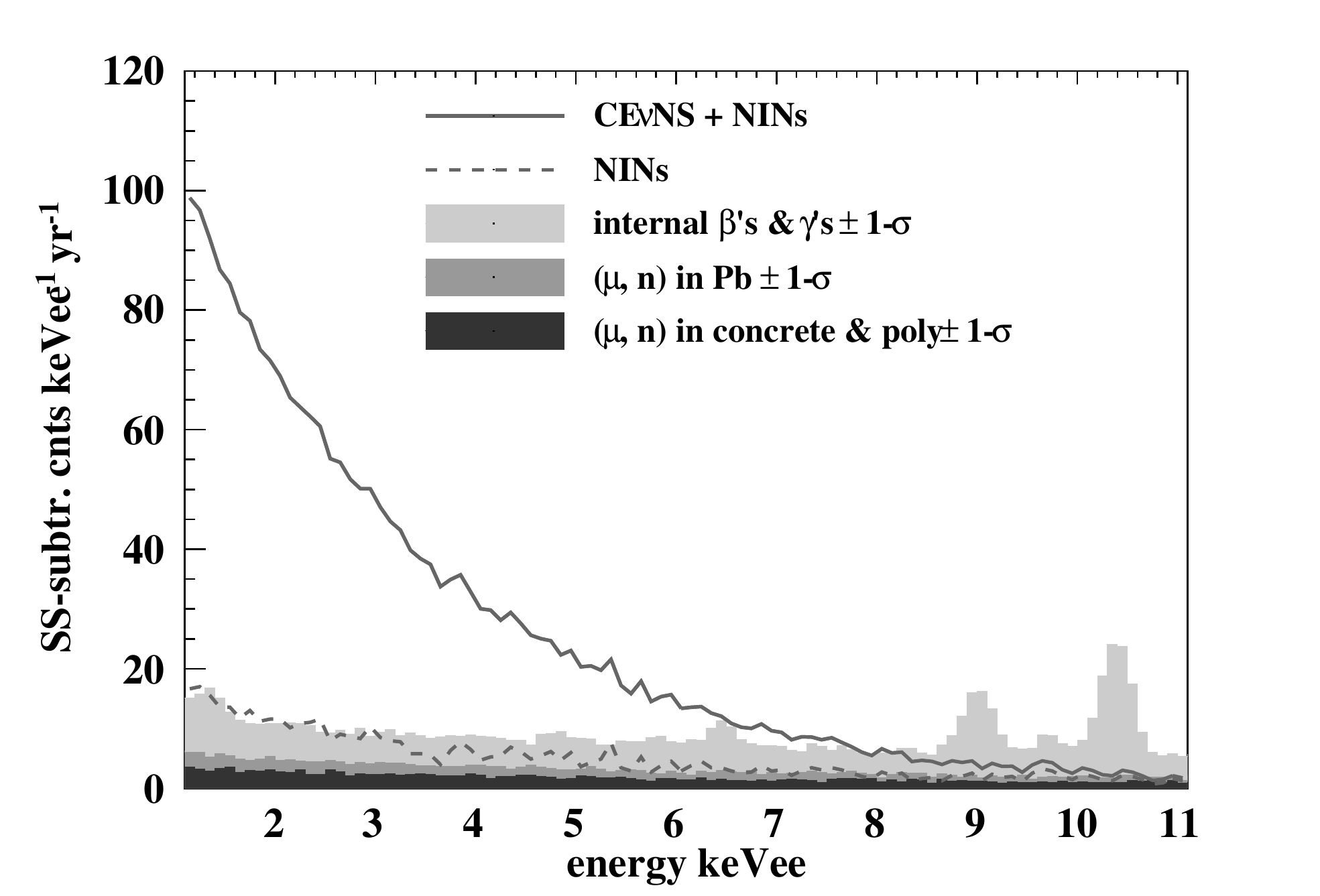}
\label{fig:spectrumge}}
\quad
\subfigure[Xe]{%
\includegraphics[height=2.0in]{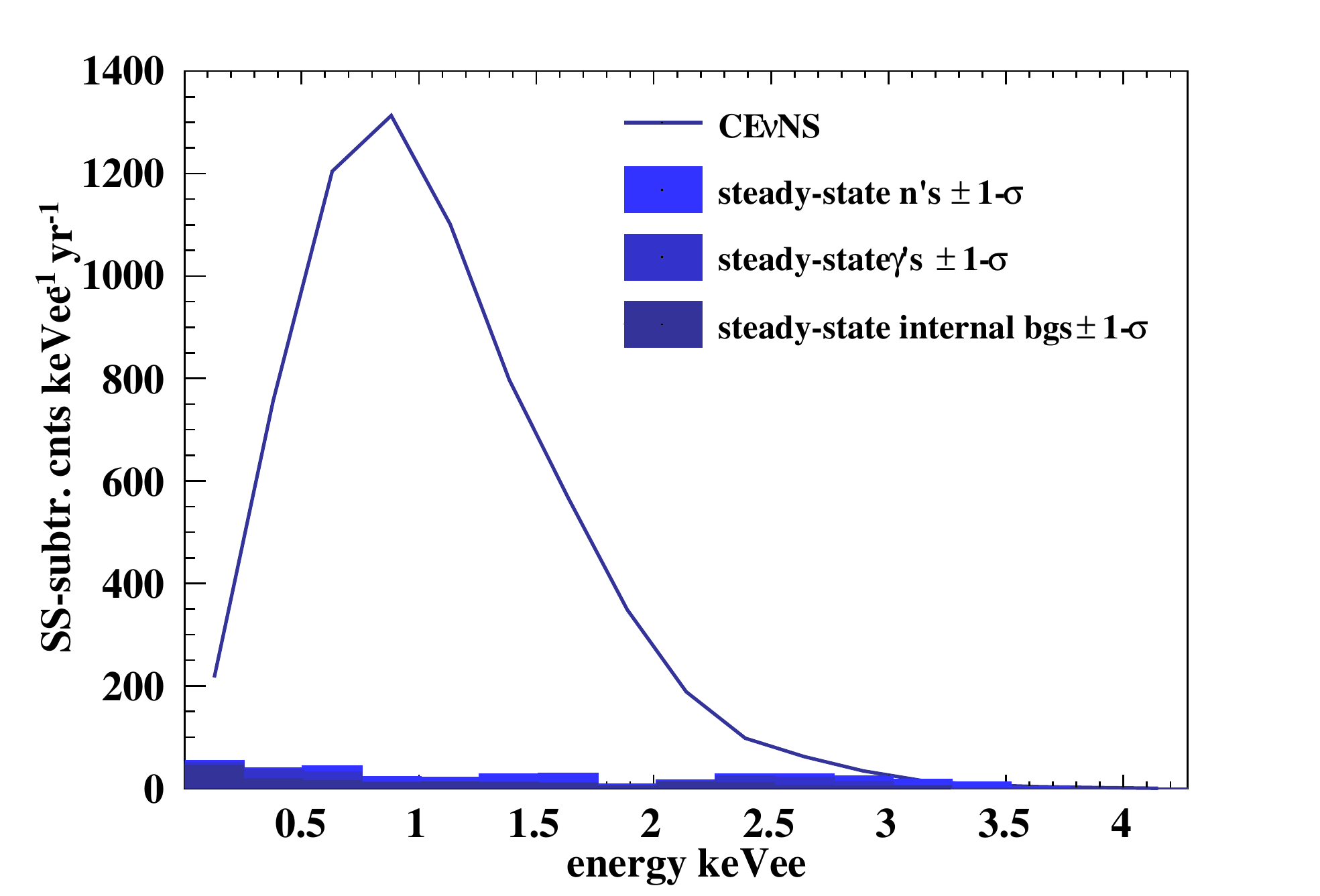}
\label{fig:spectrumxe}}
\caption{ Steady-state (SS) background-subtracted NINs background (dashed lines) with CEvNS signal (solid lines) for each detector subsystem are compared to the 1-$\sigma$ statistical fluctuations of the SS-background (shaded). Detector threshold effects and analysis timing cuts are accounted for. (a)~CsI[Na] (b)~Ge (c)~Xe. 
With 15 cm of water-based neutron shield between the LXe detector and its lead shield, NINs are expected to contribute only $\sim$25 events per year.}\label{fig:sigplusbg}
\end{figure}

\input{Ge.tex}

\input{Xe.tex}

%% file: CsI.tex
\subsubsection{CsI[Na] Detector Subsystem}
\paragraph{CsI[Na] Scintillator Detectors for CEvNS}

CsI[Na] scintillators (cesium iodide crystals doped with sodium) present several advantages for a CEvNS measurement. This mature technology combines sufficiently low thresholds with large neutron numbers ($N=74, 78$) that make an observation of this process at the SNS feasible. Both recoiling species are essentially indistinguishable due to their very similar mass, greatly simplifying understanding the detector response.  CsI is a rugged, room-temperature detector material, and is also relatively inexpensive ($\sim$ \$1/g).
These detectors have several other practical advantages. CsI[Na] exhibits a high light yield (64 photons/keVee (electron-equivalent energy deposition)) and has the best match to the response curve of bialkali photomultipliers of any scintillator material. CsI[Na] also lacks the excessive afterglow (phosphorescence) that is characteristic of CsI[Tl]~\cite{cosinima}, an important feature in a search involving small scintillation signals in a detector operated at ground level. The quenching factor for nuclear recoils (the fraction of the recoil energy that is detectable as scintillation) in this material over the energy region of interest has been carefully characterized~\cite{cosinima}, using the methods described in~\cite{Collar:2013gu}. The quenching factor is sufficiently large to expect a realistic $\sim$5 keVnr (nuclear-recoil energy deposition) threshold. Additional quenching factor measurements are planned by the Collaboration using monochromatic neutron beams, to ensure the minimization of systematic uncertainties. Background studies in conditions similar to those at the SNS have been performed~\cite{cosinima}, confirming the suitability of this detector technology.

Neutron background measurements were performed in 2014 at the intended SNS basement location for this CsI[Na] detector, with the shield populated by two liquid-scintillator neutron detectors representing a similar detector volume to a 14-kg CsI[Na] crystal. A nuclear recoil rate of only $\sim$1 per day was measured in coincidence with the SNS trigger signal. These events are restricted to a very narrow time region, which bodes well for the identification of prompt CEvNS signals. A second conclusion from these background studies is that events induced by neutron emission following $^{208}\mathrm{Pb}(\nu_e,e^-)^{208}\mathrm{Bi}$ (NINs) in a lead shield can be sufficiently reduced by the addition of an internal layer of high-density polyethylene (HDPE) (Fig.~\ref{fig:csi_installation}) for neutron absorption.

A 14-kg CsI[Na] detector and shielding (Fig.~\ref{fig:csi_installation}) has been completed and was installed at the SNS in June 2015. 
The shielding consists of 7.5 cm of HDPE inner neutron moderator, 15 cm of Pb (the innermost 5 cm selected for low $^{210}$Pb content), a 99.6\%-efficient muon veto, and an outer neutron moderator and absorber.
See reference~\cite{cosinima} for more details of the detector and its characterization.

The large CEvNS cross section from both recoiling species, Cs and I, generates $\sim$400 recoils per 14 kg of CsI[Na] per year above the expected $\sim$5 keVnr threshold of this detector, that also survive the 1-$\mu$s window timing cut selecting delayed neutrinos.  This should result in a 5$\sigma$ measurement of CEvNS in two years of running time (one year running time defined for the purposes of this proposal as 365 days of 1.4-MW SNS operation).  Other than the 10\% neutrino flux uncertainty common to all three detectors, the dominant systematic uncertainty for the CEvNS signal in CsI[Na] is due to the 10\% uncertainty of the quenching factor at threshold, which results in a 7\% uncertainty on the expected interaction rate above threshold for delayed neutrinos (Fig.~\ref{fig:cross-sections}). Figure~\ref{fig:spectrumcsi} shows the expected signal spectrum and fluctuations of background for a year of running time.

%% file: Ge.tex
\subsubsection{P-Type Point Contact HPGe Subsystem}
\paragraph{P-Type Point Contact Germanium Detectors for CEvNS}

P-type point-contact (PPC) High-Purity Germanium (HPGe) detectors are well matched to the problem of detecting CEvNS-induced recoils.   They combine low thresholds, high energy resolution, an adequate (and well-understood) quenching factor~\cite{Aalseth:2012if} and low internal backgrounds.  
The moderate atomic number ($72 \le A \le 76$) also leads to more energetic coherent recoil scatters than from heavier nuclei.

Point-contact HPGe detectors have very small electrodes, and so have very small detector capacitance, and consequently very low electronic noise~\cite{Luke:1989}.  When cooled, the leakage currents can also be very low --- less than 1 pA --- so the current noise is also small.  PPC detectors with an electronic noise FWHM of order 160~eV have been demonstrated~\cite{Aalseth:2011wp}, and a 1-keVee threshold has been routinely achieved~\cite{Aalseth:2011wp,Aalseth:2012if,Barbeau:2009zz,lin2009:texono}.

In addition to extremely low energy thresholds, the low noise gives these detectors an excellent energy resolution. As a result the measured (background-subtracted) energy-deposition spectrum is faithful to the true recoil spectrum. This allows straightforward searches for deviations of the recoil spectrum due to nuclear form factors.  Additionally, there are several well-resolved peaks that result from the cosmogenic activation of electron-capture decaying isotopes in the germanium; these steady-state lines are useful calibration points which can be largely subtracted from the final data.
 
The high intrinsic radiopurity of the germanium and the availability of low-background cryostats makes low-background operations possible; existing PPC detector deployments have demonstrated intrinsic backgrounds sufficiently low for CEvNS detection~\cite{Aalseth:2012if,Giovanetti:2014fhx}.  The 1-$\sigma$ fluctuations from an energy spectrum obtained with the MALBEK detector (see~\cite{Giovanetti:2014fhx}) represent internal backgrounds in Fig.~\ref{fig:spectrumge}.   Additionally, the drift times in PPC detectors are
of order 1 $\mu$s~\cite{Martin:2011vj}; at the SNS, this is adequate to separate background neutrons coincident with the 800-ns full-width beam pulse and the delayed neutrinos from muon decay.

\paragraph{Design overview}
The COHERENT Collaboration has the ability to directly leverage germanium detector array technology developed by the dark-matter and double-beta decay communities.  
In particular, the \mjd~(MJD) double-beta decay experiment has been designed for deployment of PPC germanium detectors in modular ultra-low-background vacuum cryostats in a compact shielding configuration~\cite{Abgrall:2014}.  Each of the \sc Demonstrator's \rm two vacuum cryostats has the capacity to support up to $\sim$22~kg of detectors; a single MJD~module surrounded by a compact copper, lead, and polyethylene shield is a deployment configuration suitable for a CEvNS measurement, and one that can be rapidly implemented with minimal engineering effort. 

In the process of developing the MJD, a Prototype Module was assembled
from materials which do not meet the strict radiopurity requirements of the \sc Demonstrator. \rm The \sc Majorana \rm Collaboration Executive Committee (MEC) has agreed to allow the COHERENT Collaboration use of the MJD Prototype Module's cryostat, internal cryogenic hardware, and support frame.  The MEC has also agreed to allow COHERENT to use eight unenriched PPC detectors ($\sim$5~kg total) unneeded for the \sc Demonstrator \rm as targets for a CEvNS measurement, along with spare mounting parts and the CNC milling machine code and fixturing required to produce additional hardware. 

In addition to the 5~kg of detectors promised by \mj, Los Alamos National Laboratory (LANL) is in possession of an additional 5~kg of PPC detectors which can be used for COHERENT.  Finally, the COHERENT Collaboration will purchase 5~kg of additional PPC detectors for the deployment of a 15-kg target, in an available cryostat, using components already designed and tested.  COHERENT will also take advantage of the significant development work which \mj~has invested in data acquisition hardware and software, plus simulations and analysis tools. 

\paragraph{Cryogenic Module and Detectors}
The MJD Modules are vacuum and cryogenic platforms for deployment of arrays of germanium detectors.  COHERENT will construct a single MJD Module around the Prototype Module Cryostat from \sc Majorana. \rm  A new vacuum system will be constructed based on the MJD design for use with the cryostat.  Cooling of the detector array will be provided by a liquid-nitrogen-(LN$_2$)-driven thermosyphon system \cite{Aguayo:2013}.  
This thermosyphon-based system requires only regular, automated LN$_2$ refills for stable operation, and causes significantly less noise-inducing vibration on the detector array than direct LN$_2$ cooling by cold-finger. 

A total of $\sim25$ detectors comprising 15~kg of target mass will be deployed.  To keep the detectors free of background-causing surface contamination and prevent leakage-current-inducing damage to the passivated surfaces of the detectors, they will be handled and installed within a nitrogen-purged glovebox. 
The PPC detectors will be installed in low-background copper mounts of the design developed by \sc Majorana, \rm then stacked into strings of 5 detectors each for installation in the Module. 

\paragraph{Deployment Location \& Shielding}

\begin{figure}[ht]
\centering
\subfigure[Side view, toward SNS target]{%
\includegraphics[width=0.45\linewidth]{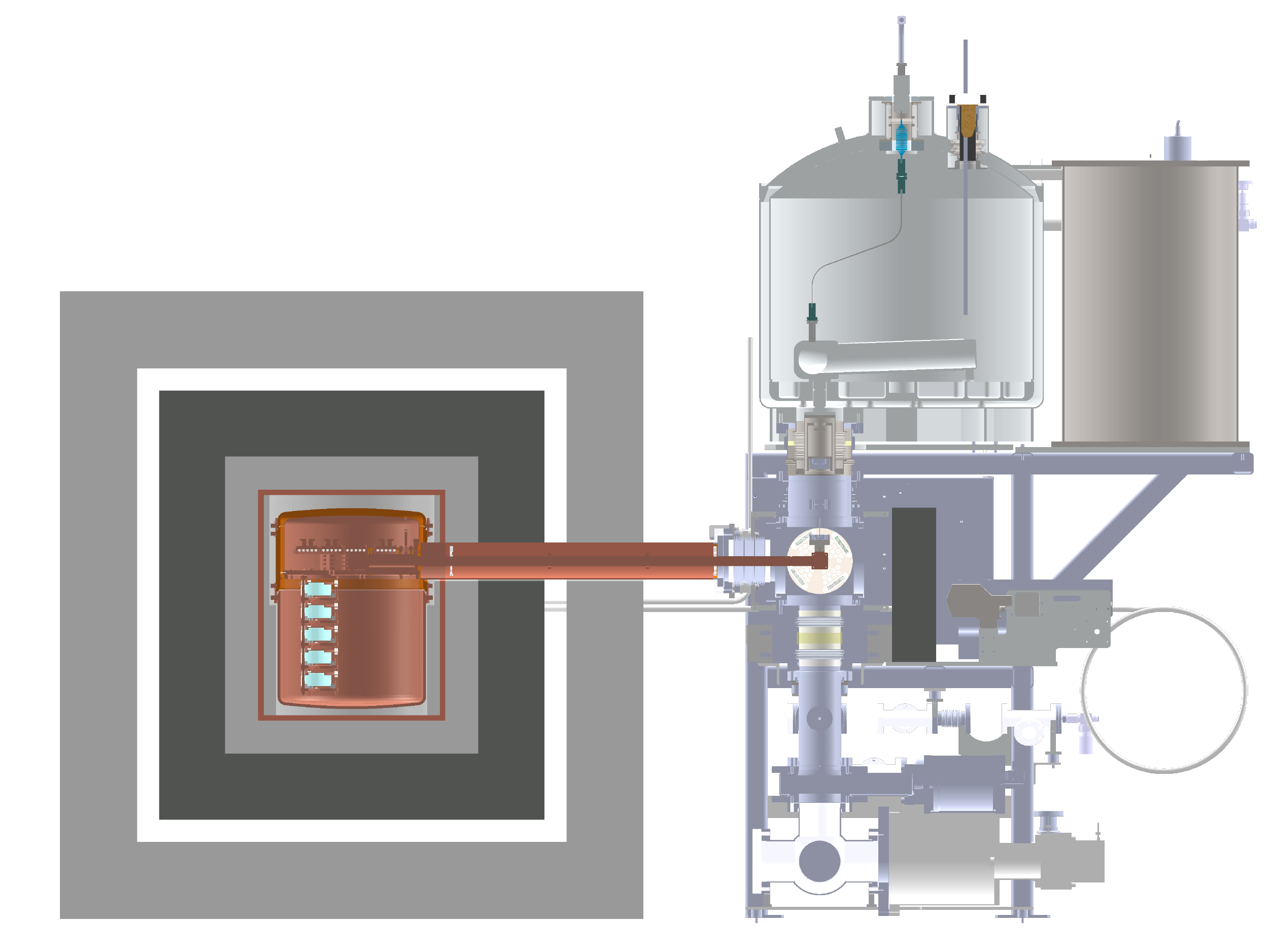}
\label{fig:GeShieldingSide}}
\quad
\subfigure[Top view]{%
\includegraphics[width=0.45\linewidth]{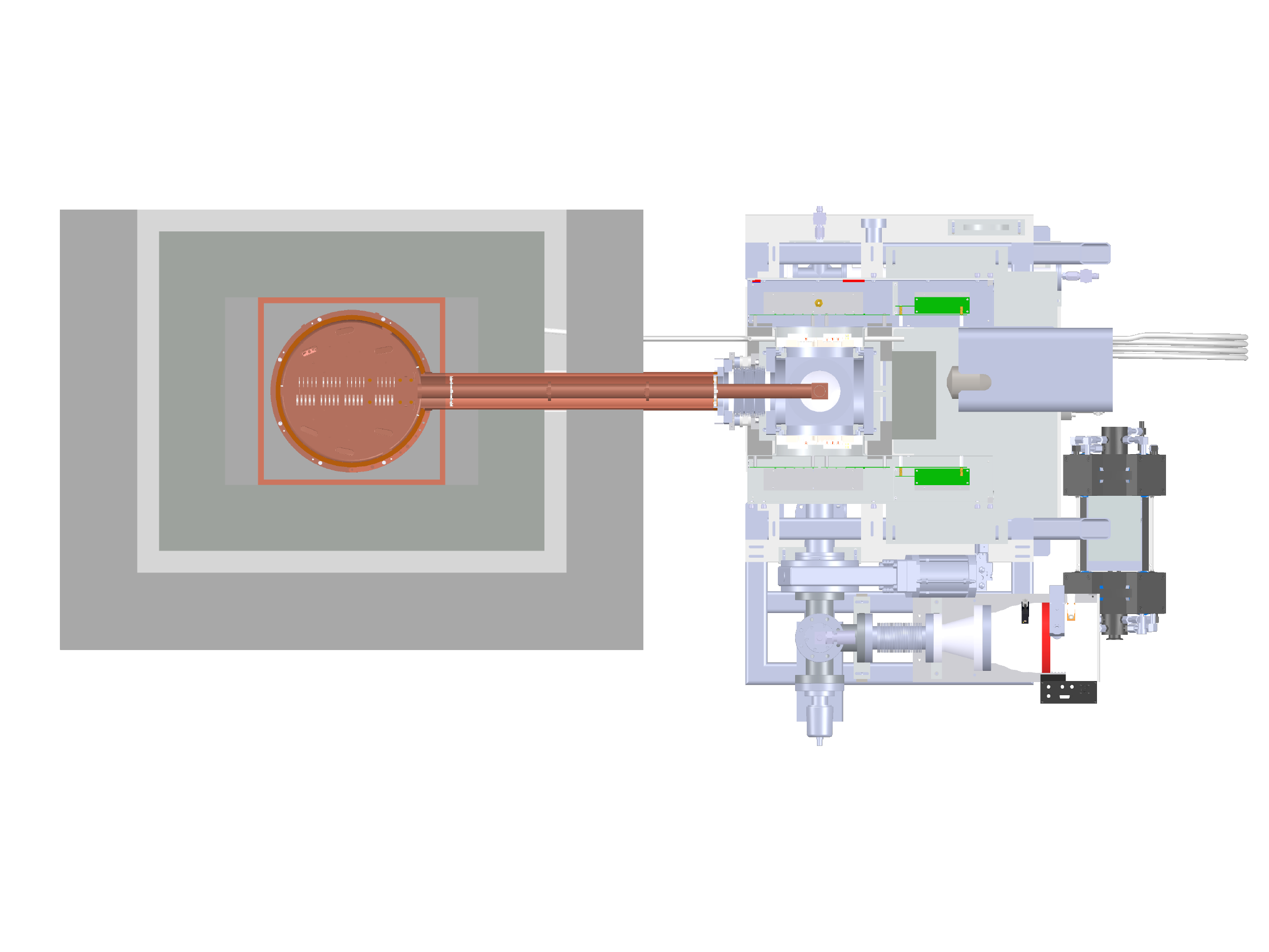}
\label{fig:GeShieldingTop}}
\caption{Shielding configuration of the germanium detector array subsystem.  The concrete shielding wall is used to provide shielding from low-energy neutrons in the hallway, thus the asymmetric neutron shielding configuration.}
\label{fig:GeShielding}
\end{figure}

The background measurements described in Sec.~\ref{sec:backgroundstudies} have made clear that the basement hallway provides the best combination of neutrino source proximity, fast neutron shielding, and low environmental backgrounds in the SNS target building.  The germanium detector subsystem will be sited in that hallway, against the wall closest to the SNS target, and adjacent to the existing CsI subsystem deployment.  The detector array will be surrounded by a layered radiological shield as shown in Fig.~\ref{fig:GeShielding}, with an asymmetric design constrained by personnel egress requirements for the basement hallway.  A 5-sided, 18-cm-thick polyethylene shield comprises the outermost layer of shielding, for the absorption of slow neutrons which permeate the hallway.  A 5-cm-thick plastic scintillator $\mu$-veto is nested inside of the poly, and surrounds a 15-cm-thick lead $\gamma$-ray shield.  To limit the NIN background, a 4-sided, 7.5-cm-thick polyethylene shield is placed inside the lead to provide shielding in the directions where additional material does not interfere with egress requirements.  A 12.5-mm-thick copper box shields bremsstrahlung from $^{210}$Pb decays, and additional polyethylene is inserted between the copper box and cryostat for further NINs background reduction.  MCNPX-PoliMi~\cite{pozzi:2012} simulations have demonstrated the effectiveness of this shield design, as seen in Fig.~\ref{fig:spectrumge}; further simulations will be performed to optimize it.

\paragraph{Data Acquisition (DAQ)}
The charge read-out scheme in use by \mj~was developed to enable low detector thresholds for a competitive light-WIMP search, and background reduction from $^{68}$Ge through detection of time-correlated 1-keV Ga x-rays.  These goals are in line with establishing an energy threshold suitable for CEvNS~detection at SNS.  Charge is read out from the germanium detectors through a low-noise, low-background, resistive feedback front-end.  
Preamplifiers located outside of the shielding amplify the signals for digitization and Struck 3302 waveform digitizers acquire waveforms from each detector, triggered by the SNS timing signal.  Off-line analysis of the acquired waveforms may allow lower energy thresholds than standard trapezoidal or leading-edge discriminating triggers. We will use ORCA (Object-oriented Real-time Control and Acquisition)~\cite{Howe:04} DAQ software, which has already been fully developed for \mj{}, including analysis software and a control application for the Module systems. Only minor software modifications will be required for COHERENT.

\paragraph{CEvNS background and detection sensitivity}
Approximately 220 CEvNS events per year are expected for neutrinos originating from $\mu$-decay in the SNS target (delayed neutrinos) that pass an energy threshold of 1~keVee and also survive a timing cut intended to eliminate beam-coincident fast neutron backgrounds. Backgrounds from radioactivity internal to the detector array, external gamma-ray sources, environmental and $\mu$-induced neutrons and NINs have been considered.  Of these, all but the NINs are reduced by a factor of $\sim1000$, and can be measured to high accuracy in the time period between neutrino pulses; only the statistical fluctuations of those sources are backgrounds for CEvNS.  Internal backgrounds have been estimated in the CEvNS region of interest between 1-10~keVee based on the background observed in the MALBEK detector.  Unpublished measurements of the backgrounds measured in the MJD Prototype Module support this scaling argument.  Gamma-ray backgrounds are estimated based on background measurements performed at the experiment location in the CsI shield.  MCNPX-PoliMi Monte Carlo simulations have been performed to assess backgrounds from environmental and $\mu$-induced neutrons.  Unlike the steady-state backgrounds, NINs arrive with an identical time structure to that of the CEvNS signal, and must be quantified through measurement of the cross section with the Neutrino Cubes described in Sec.~\ref{sec:backgroundstudies} and Monte Carlo simulations of the shielding geometry.  Preliminary simulations of the shielding model yield the NINs spectrum indicated in Fig.~\ref{fig:spectrumge}.  A 5$\sigma$ measurement of CEvNS is expected in a few months of operation. The dominant systematic uncertainty that is not due to the uncertainty of the neutrino flux (correlated between detectors) is due to the 10\% uncertainty in the quenching factor at threshold, which results in a 2\% uncertainty in the interaction rate above threshold for delayed neutrinos (Fig.~\ref{fig:cross-sections}). Figure~\ref{fig:spectrumge} shows the expected signal spectrum and fluctuations of background for a year of running time.

%% file: Xe.tex
\subsubsection{Two-phase Xenon Subsystem}

\paragraph{Two-phase Xenon Detector for CEvNS}
The deployment of a two-phase ``wall-less'' xenon emission detector to the SNS as part of a measurement of CEvNS is a natural choice considering the predominant role the technology plays in the search for dark matter~\cite{akerib2014:luxResults}. The emission method of particle detection allows measurement of single ionization electrons, generated in massive non-polar dielectrics such as condensed noble gases.  Low thresholds for nuclear recoils are achievable when the ionization electrons are extracted from the liquid phase and produce electroluminescence in the gas phase. Both the ionization electrons and primary scintillation photons can be used to reconstruct the energy deposited in the LXe, the position, and the time of particle interactions, and also provide particle identification~\cite{Bolozdynya:1995}. 

The principle of operation of a two-phase ``wall-less'' detector is more involved than both the CsI[Na] and PPC germanium detectors. Radiation interacts with the LXe, exciting and ionizing atoms. A prompt signal is generated that manifests itself in the form of scintillation. Detection of this signal by photodetectors serves as a trigger, yielding the time information of the primary interaction. In response to an applied external electric field, the ionization electrons drift to the surface of the condensed medium where they pass through the surface potential barrier into the equilibrium gas phase, where they generate a secondary scintillation signal due to the electroluminescence in a strong electric field. An array of photodetectors determine the x-y coordinate of the event by measuring the two-dimensional distribution of the secondary photons. The z-position of the event is identified by the time separation between the primary and secondary scintillations. 

The three-dimensional position information allows the definition of a fiducial volume, so that events originating in the vicinity of the detector walls and associated with radioactive backgrounds from surrounding materials can be rejected. This active shielding of the fiducial volume allows the rejection of events associated with multiple scattering of background particles. Furthermore, analysis of the deposited energy distribution between the ionization and scintillation signals provides particle identification, further suppressing backgrounds.

\begin{figure}[ht]
\centering
\subfigure[]{%
\includegraphics[width=0.25\linewidth]{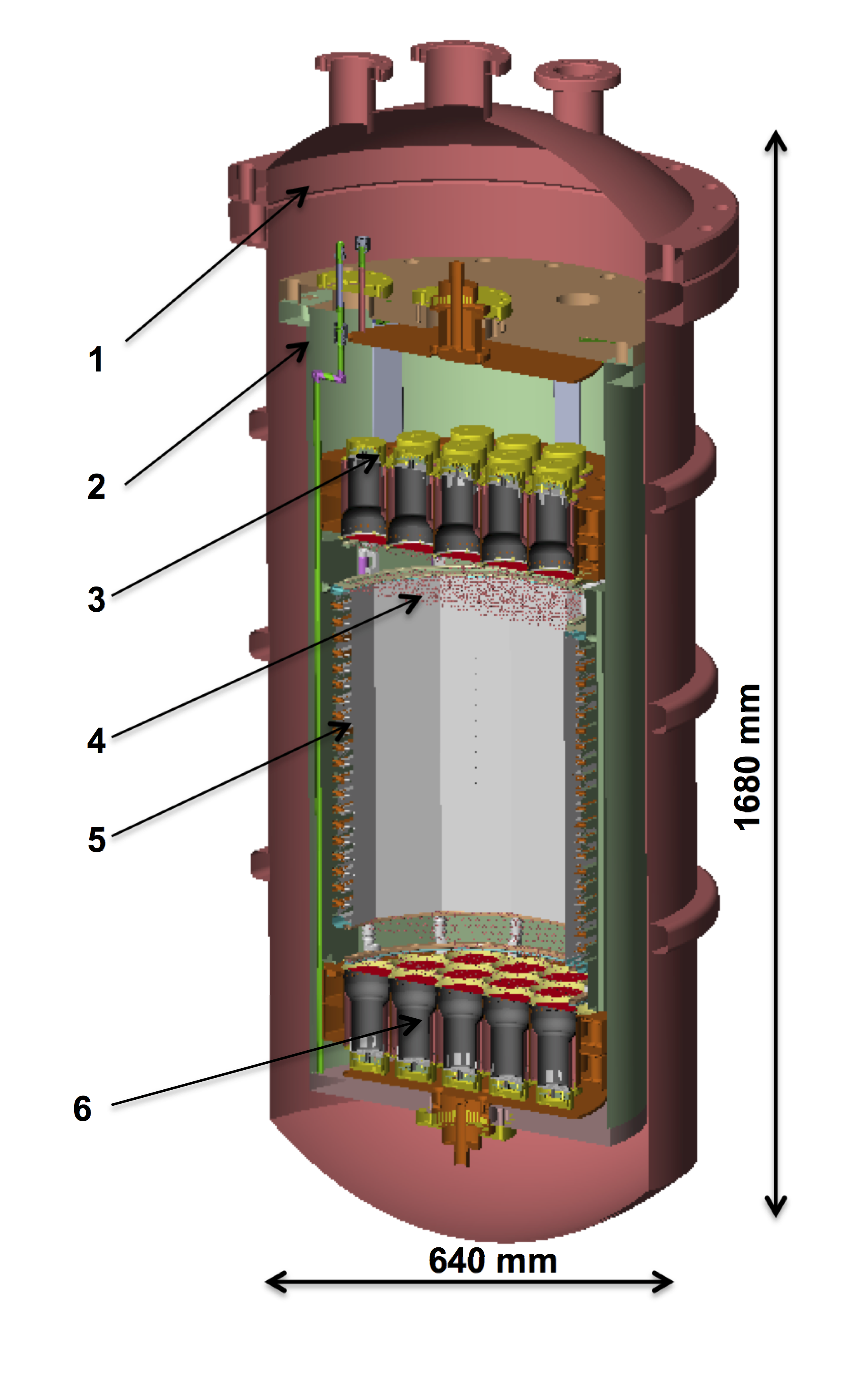}
\label{fig:Red100}}
\subfigure[]{%
\includegraphics[width=0.5\linewidth]{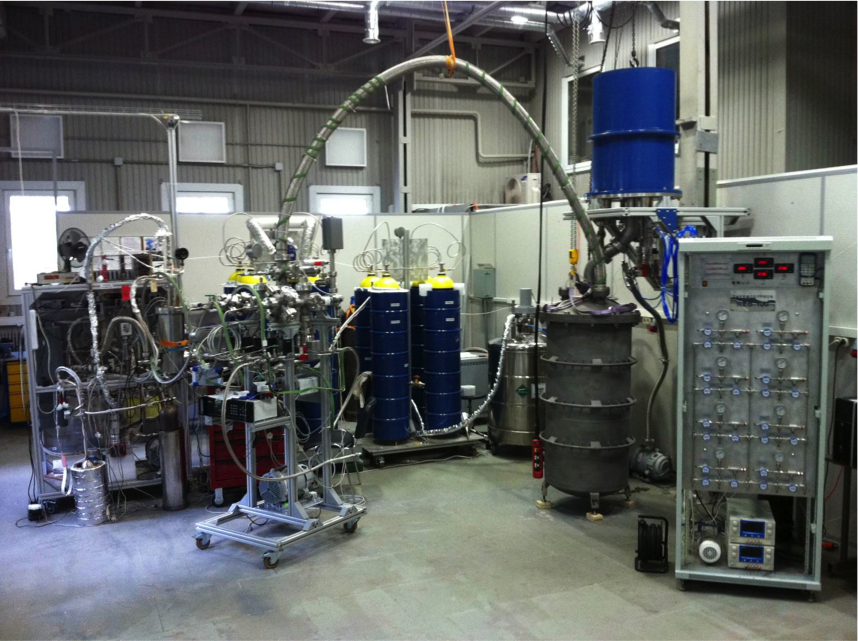}
\label{fig:RedApparatus}}
\caption{(a)~Engineering drawing of RED-100 xenon assembly: (1) outer (warm) Ti vessel (2) inner (cold) Ti vessel with Cu liner (3) top array of 19 photomultipliers (4) gridded electrodes (5) drift cage with Teflon$^{\rm TM}$ light-collection reflectors (6) bottom array of 19 photomultipliers. (b)~Detector assembly during the technical run in June 2015.}
\label{fig:Red-Overall}
\end{figure}

\paragraph{Design Overview}
The overall design and initial testing of the xenon detector assembly took place at the National Research Nuclear University in the Moscow Engineering Physics Institute (MEPhI). The detector is enclosed in a titanium cryostat as shown in Figure~\ref{fig:Red100}. Inside the cryostat there is a drift cage filled with liquid noble gas surrounded by reflective Teflon$^{\rm TM}$ panels.  Either liquid xenon and liquid argon can be used as a working medium in the detector (although COHERENT plans to use xenon). The volume of the drift cage is viewed from top and bottom by two arrays of 19 3''-diameter Hamamatsu R11410-20 photomultiplier tubes (PMTs). A quasi-uniform electric field inside the cage is formed with gridded electrodes on the top and the bottom and drift electrodes embedded into the Teflon$^{\rm TM}$ liner.  The low-background PMTs are specially designed to work in LXe at low temperatures and demonstrate a quantum efficiency of about 30\% at the 175-nm wavelength of xenon emission photons.  Due to the 1-cm wide electroluminescence gap, the detector is sensitive to single-electron ionization. To extend their longevity, the photomultipliers are outfitted with specially designed active bases to suppress PMT operation when muons pass through the detector~\cite{Akimov:2014mua}.

\paragraph{Deployment Plan} 
The experimental setup for operation of the detector consists of the detector, cryogenics, gas storage and purification equipment, and interfaces.  The experimental assembly is shown in Fig.~\ref{fig:RedApparatus}. 
The xenon detector system will be delivered from MEPhI with all of the detector components. The detector and subsystem will initially be assembled in the temporary staging area in the SNS building that has been provided to the COHERENT collaboration. Engineering and detector commissioning will take place in this location. Volume calibration will be achieved using a gaseous $^{83m}$Kr source produced according to methods previously demonstrated~\cite{bodine:phd2015}.

\paragraph{Deployment Plan: Deployment Location \& Shielding} 
After the energy resolution, light collection and electron lifetime have been verified to be within specifications, the detector will be installed at the SNS in the basement of the experimental building at a location 29~m from the target. 
A full GEANT4 simulation of the xenon apparatus has been run. Backgrounds from the radioactivity of internal components, external gammas and neutrons from the SNS were all considered.  Optical tracking of scintillation and luminescence light was also performed. The optimal shielding configuration is 15 cm of water surrounding entire detector followed by 10-15 cm of lead. The detector is designed to be submerged into a prefabricated water tank in order to provide shielding against neutrons. In addition lead placed outside the water tank will shield the detector from gamma backgrounds. There is some optimized non-uniformity of the required lead thickness due to the non-isotropic flux of gamma backgrounds in the basement.

\paragraph{CEvNS background and detection sensitivity}
Approximately 1,700 CEvNS events per year are expected for neutrinos originating from $\mu$-decay in the SNS target (delayed neutrinos), with sufficient energy to provide a prompt scintillation trigger and that also survive the 1 $\mu$s timing cut intended to eliminate beam-coincident fast neutron backgrounds.  The background simulation indicates that the expected background is an order of magnitude less than the signal from CEvNS. A 5$\sigma$ measurement of CEvNS is expected within a few months of operation.  The dominant systematic uncertainty that is not due to the uncertainty of the neutrino flux (correlated between detectors) is due to the 18.9\% uncertainty in the quenching factor at threshold~\cite{mock:2014NEST}, which results in a 13\% uncertainty in the interaction rate above threshold for delayed neutrinos (Fig.~\ref{fig:cross-sections}). Figure~\ref{fig:spectrumxe} shows the expected signal spectrum and fluctuations of background for a year of running time.

%% file: timeline.tex
\section{Status and Timeline}

\begin{figure}[ht]
\centering
\hspace*{\fill}
\includegraphics[height=6.0cm]{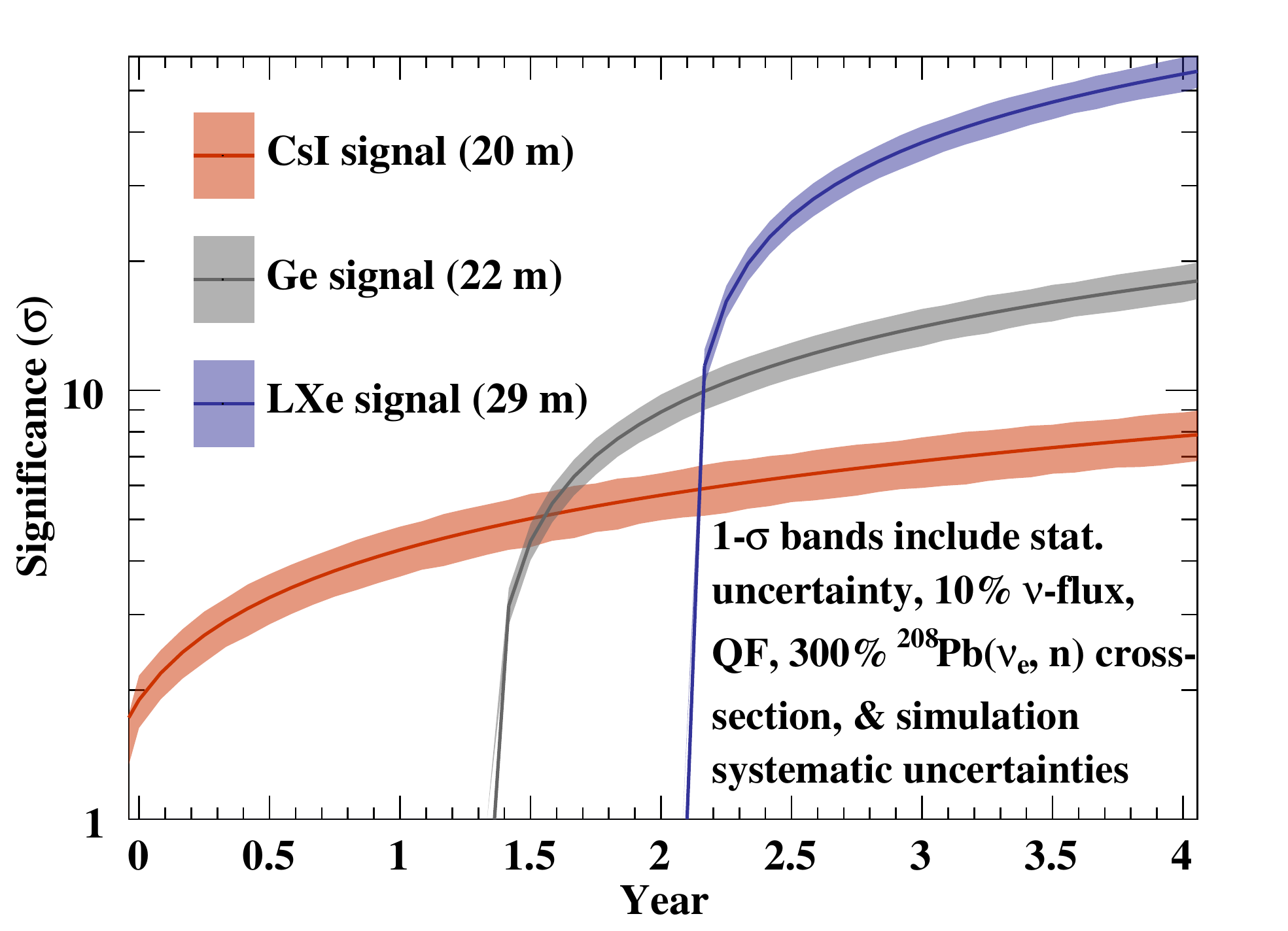}
\hspace*{\fill}
\caption{Sensitivity versus running time. The 1-$\sigma$ bands reflect the statistical fluctuations of the signal and backgrounds as well as the effect of the systematic uncertainties in the interaction rate due to imprecise knowledge of the quenching factors at threshold, the 10\% neutrino flux uncertainty, the Pb($\nu_e$, n) production rate and simulation efficiencies.}
\label{fig:sens}
\end{figure}

The anticipated sensitivity as a function of time for COHERENT is shown in Fig.~\ref{fig:sens}.  The status of the three detector subsystems as of this writing (September 2015) is as follows:

\begin{description} 
  \item[CsI] The detector has been installed and is already operational.  Approximately three years of data taking will be required to achieve the desired sensitivity.
  
  \item[Ge] Most of the detectors, as well as the cryostat, exist and are available for shipment to ORNL once funding has been established.  The purchase of additional germanium detectors to supplement those already supplied will be a long-lead-time item and will be initiated as soon as funding has been established.  Front-end electronics fabrication will take approximately a year before final detector assembly and characterization can be performed.   The designs for the external active/passive shielding will be finalized concurrent with the front-end development.  Final construction of the detector system, shielding system and installation in the SNS basement are expected during the second year of the project.  Two years of data taking are required to achieve the eventual sensitivity.
  
  \item[Xe] The two-phase xenon detector exists and has undergone testing in Russia.  An FY16 ORNL Laboratory Directed Research and Development (LDRD) has been approved to install, shield, and commission the system by the end of FY17. Once the detector has arrived at ORNL (in the first quarter of FY2016), approximately one year of Xe purification will be required before final installation and commissioning can begin.  During that time, the design and construction of the external active/passive shielding will be performed.  In addition, the Xe deployment will require more extensive interfacing with the SNS to provide for necessary ancillary components. Because the Xe system has the largest mass and therefore the highest statistical power, it will also have the largest margin of error for schedule delays, requiring only 3-6 months to achieve the desired sensitivity.

\end{description}

%% file: summary.tex
\section{Summary}

In summary, the COHERENT program will take advantage of the extremely high-quality stopped-pion neutrino source available at the Spallation Neutron Source at Oak Ridge National Laboratory for new CEvNS measurements.
Three detector subsystems, 
based on  CsI, Ge, and Xe, will measure CEvNS at  5$\sigma$ significance, and
to demonstrate the $N^2$ dependence of the cross section. 
A secondary goal of COHERENT is to measure the cross sections for
NINs on lead, iron, and copper using independent scintillator detectors, in
order to estimate the NIN component of the CEvNS detector signal
sufficiently well to achieve the primary goal. These measurements will be interesting physics results in
their own right.
  Beyond the planned four-year program, there is physics motivation (see Sec.~\ref{sec:future}) for several potential upgrades, including increased mass for any of the existing
  targets, and additional targets such as NaI and argon.  
